\documentclass[10pt, journal]{IEEEtran}
\IEEEoverridecommandlockouts
\usepackage{url}
\usepackage{cite}
\usepackage{flushend}
\usepackage{makecell}
\usepackage{amsmath,amssymb,amsfonts,commath}
\usepackage{amsthm}
\usepackage[version=4]{mhchem}
\usepackage{graphicx}
\usepackage{textcomp}
\usepackage{float}
\usepackage{multirow}
\usepackage[normalem]{ulem}
\useunder{\uline}{\ul}{}
\usepackage[table,xcdraw]{xcolor}
\usepackage{booktabs}
\usepackage{threeparttable}
\usepackage[caption=false]{subfig}
\usepackage{wasysym}

\usepackage[utf8]{inputenc}
\usepackage{algorithmic}
\usepackage[linesnumbered,ruled]{algorithm2e} 

\renewcommand{\algorithmiccomment}[1]{\hfill $\triangleright$ #1}

\newcommand{\BandWidthSingle}{b_{ji}}
\newcommand{\CarrierFrequency}{f_c}
\newcommand{\TransPower}{p_{ji}^{\text{tx}}}
\newcommand{\TransNoise}{p_{ji}^{\text{noise}}}
\newcommand{\PathLoss}{p_{ji}^{\text{loss}}}
\newcommand{\ThresholdCollab}{p_{\text{thre}}}

\newcommand{\LenDrivingMemory}{T_d}
\newcommand{\LenDrivingTrajectory}{T_f}
\newcommand{\Height}{H}
\newcommand{\Width}{W}
\newcommand{\Channel}{D}
\newcommand{\NumClass}{C}
\newcommand{\FocusRadius}{\sigma_F}

\newcommand{\DiscountedRate}{\gamma}
\newcommand{\WeightLossVgg}{\varepsilon}
\newcommand{\NoisePosition}{\sigma_p}
\newcommand{\NoiseRotation}{\sigma_r}
\newcommand{\NAgents}{\mathcal{N}_i^t}
\newcommand{\AllAgents}{i\cup\NAgents}
\newcommand{\SensorEgo}{X_i^t}

\newcommand{\PseuEgo}{\mathcal F_{i}^t}
\newcommand{\PseuBack}{\mathcal F_{j}^t}

\newcommand{\PseuBackToPredict}{\mathcal F_{j}^{\ReceiveTime}}
\newcommand{\PseuBackProcessed}{\widehat{\mathcal F}_{j}^{\ReceiveTime}}
\newcommand{\PseuFused}{\widetilde{\mathcal F}_{i}^{\ReceiveTime}}

\newcommand{\BoundingBoxEgo}{\mathcal B_{i}^t}

\newcommand{\BoundingBoxFused}{\widetilde{\mathcal B}_{i}^{\ReceiveTime}}
\newcommand{\BoundingBoxGT}{{\overline{\mathcal B}_i^t}}

\newcommand{\HeatMapEgo}{\mathcal H_{i}^t}
\newcommand{\HeatMapBack}{\mathcal H_{j}^t}
\newcommand{\HeatMapBackToPredict}{\mathcal H_{j}^{\ReceiveTime}}

\newcommand{\HeatMapBackNext}{\mathcal H_{j}^{t+\IntervalOverall}}

\newcommand{\HeatMapBackHistoryBatchAll}{\HeatMapProcedureGT{t-\IntervalOverall
}, \HeatMapProcedureGT{t}}
\newcommand{\HeatMapProcedure}[1]{\widehat{\mathcal H}_{j}^{#1}}

\newcommand{\HeatMapProcedureGT}[1]{\mathcal H_{j}^{#1}}
\newcommand{\HeatMapFused}{\widetilde{\mathcal H}_{i}^{\ReceiveTime}}
\newcommand{\HeatMapGT}{\overline{\mathcal H}_i^t}
\newcommand{\HeatMapGTReceive}{\overline{\mathcal H}_i^{\ReceiveTime}}

\newcommand{\HeatMapAPCGT}[2]{\overline{\mathcal H}_{#1}^{#2}}
\newcommand{\Temperature}{\beta}

\newcommand{\GumbelNoise}{G_k}
\newcommand{\expfunc}[1]{\text{exp}\left( \dfrac{1}{\Temperature} ( #1 )\right)}

\newcommand{\PassingMask}{\mathcal P_{ji}^t}
\newcommand{\VoxelFlow}[1]{\mathcal V_{j}^{#1}}
\newcommand{\RoutingVector}{v^{k}}
\newcommand{\RoutingProb}{\tilde{v}^k}
\newcommand{\ScalingFactor}{S^{k}}
\newcommand{\ConfidenceMapEgo}{\mathcal C_{i}^t}
\newcommand{\ConfidenceMapBack}{\mathcal C_{j}^t}

\newcommand{\ConfidenceDifference}{\Delta\mathcal{C}_{j}^t}
\newcommand{\RequestMapEgo}{\mathcal R_{i}^t}

\newcommand{\Msg}{\mathcal{M}_{ji}^t}

\newcommand{\MsgSelf}{\mathcal{M}_{ii}^{t}}
\newcommand{\WeightFusion}{\mathcal{Z}_{ji}^{\ReceiveTime}}
\newcommand{\MsgAll}{\{\Msg\}_{\AllAgents}}

\newcommand{\HeatMapFlow}{\overrightarrow{\mathcal H}_j^{\ReceiveTime}} 
\newcommand{\HeatMapFlowGT}{\overrightarrow{\mathcal H}_j^{\ReceiveTime}} 
\newcommand{\OccupancyMapEgo}{\mathcal O_{i}^{\ReceiveTime}}
\newcommand{\OccupancyMapEgoHistory}{\{\mathcal O_{i}^k\}_{k=\ReceiveTime-\LenDrivingMemory}^{\ReceiveTime}}
\newcommand{\Navigation}{D_i^{\ReceiveTime}}
\newcommand{\Waypoints}{\mathcal W_i^{\ReceiveTime}}
\newcommand{\WaypointsLast}{\mathcal W_i^{\ReceiveTime-\IntervalOverall}}
\newcommand{\WaypointsGT}{\overline{\mathcal W_i}^{\ReceiveTime}}
\newcommand{\DriveAction}{\mathcal A_i^{\ReceiveTime}}

\newcommand{\ReceiveTime}{t_r}
\newcommand{\ReceiveTimeEstimate}{t_r'}

\newcommand{\LatencyOverall}{\tau_{ji}}
\newcommand{\LatencyOverallSelf}{\tau_{ii}}
\newcommand{\LatencyOverallEst}{\widetilde{\tau}_{ji}}
\newcommand{\LatencyQueue}{\tau_{ji}^{\text{q}}}
\newcommand{\LatencyTrans}{\tau_{ji}^{\text{tx}}}
\newcommand{\LatencyPropagation}{\tau_{ji}^{\text{pr}}}
\newcommand{\LatencyNetwork}{\tau_{ji}^{\text{net}}}

\newcommand{\LatencyAsync}{\tau_{ji}^{\text{asyn}}}
\newcommand{\LatencyExt}{\tau_{j}^{\text{ext}}}
\newcommand{\LatencyDecision}{\tau_{i}^{\text{dm}}}
\newcommand{\LatencyOther}{\delta \tau_{ji}}
\newcommand{\LatencyOtherEst}{\widetilde \delta \tau_{ji}} 
\newcommand{\IntervalOverall}{\tau} 
 
\newcommand{\IntervalSync}{\Delta \tau} 
\newcommand{\LatencySteps}{n_{ji}^t}
\newcommand{\Distance}{d_{ji}}
\newcommand{\EvalDrive}{\mathcal{E}}

\newcommand{\ModEncoder}{\Phi_\text{encoder}}
\newcommand{\ModProcess}{\Phi_\text{process}}
\newcommand{\ModFuse}{\Phi_\text{fuse}}
\newcommand{\ModDecoder}{\Phi_\text{decoder}}

\newcommand{\ModForecast}{\Phi_\text{DMVFN}}
\newcommand{\ModPlan}{\Phi_\text{plan}}
\newcommand{\ModController}{\Phi_\text{controller}}
\newcommand{\ModMVFB}{\Phi_\text{MVFB}^{k}}
\newcommand{\ModMAT}{\Phi_\text{MAT}}

\newcommand{\LossDMVFN}{\mathcal{L}_\text{DMVFN}}
\newcommand{\LossMAT}{\mathcal{L}_\text{MAT}}
\newcommand{\Maximum}[1]{\max\left( #1 \right)} 

\newcommand{\OneMask}[1]{\mathbf{1}\left( #1 \right)}

\newcommand{\ConfidenceMapGenerator}{\Phi_\text{Gen}}
\newcommand{\ModMHA}{\Phi_\text{MHA}}

\newcommand{\stateEstimate}{\widetilde{\mathcal{S}}_i^{\ReceiveTime}}

\newcommand{\state}{\mathcal{S}_i^{\ReceiveTime}}

\newcommand{\DistributionDataset}{\mathcal{T}}

\newcommand{\ParamPerception}{\eta}
\newcommand{\ParamPlan}{\theta}

\newcommand{\ModPerception}{\Phi_\text{percep}}
\newcommand{\PolicyExpert}{\pi_E}

\allowdisplaybreaks
\begin{document}

\title{Select2Drive: Pragmatic Communications for Real-Time Collaborative Autonomous Driving}
\author{
\IEEEauthorblockN{Jiahao Huang, \textit{Student Member, IEEE}, 
Jianhang Zhu,  \textit{Graduate Member, IEEE}, \\
Rongpeng Li,  \textit{Senior Member, IEEE}, 
Zhifeng Zhao,  \textit{Member, IEEE} and Honggang Zhang, \textit{Fellow, IEEE}, 
\thanks{Manuscript revised  July 21, 2025 and September 5, 2025; accepted September 15, 2025. This work was supported in part by the National Key Research and Development Program of China under Grant 2024YFE0200600, in part by the Zhejiang Provincial Natural Science Foundation of China under Grant LR23F010005, in part by Huawei Cooperation Project under Grant TC20240829036. Corresponding author: Rongpeng Li.} \thanks{J. Huang, J. Zhu, and R. Li are with Zhejiang University, Hangzhou 310027, China, (email: \{22331083, zhujh20, lirongpeng\}@zju.edu.cn).}\thanks{Z. Zhao is with Zhejiang Lab, Hangzhou 310012, China, as well as Zhejiang University, Hangzhou 310027, China (email: zhaozf@zhejianglab.com).} \thanks{H. Zhang is with Macau University of Science and Technology, Macau, China (email: hgzhang@must.edu.mo).}}
}

\maketitle

\begin{abstract}  
Vehicle-to-everything communications-assisted autonomous driving 
has witnessed remarkable advancements in recent years, with pragmatic communications (PragComm) emerging as a promising paradigm for real-time collaboration among vehicles and other agents. 
Simultaneously, extensive research has explored the interplay between collaborative perception and decision-making in end-to-end driving frameworks. 
In this work, we revisit the collaborative driving problem and propose the Select2Drive framework to optimize the utilization of limited computational and communication resources. 
Particularly, to mitigate cumulative latency in perception and decision-making, Select2Drive introduces distributed predictive perception 
by formulating an active prediction paradigm and simplifying high-dimensional semantic feature prediction into a computationally efficient, motion-aware reconstruction. 
Given the ``less is more" principle that an over-broadened perceptual horizon possibly confuses the decision module rather than contributing to it, Select2Drive utilizes area-of-importance-based PragComm 
to prioritize the communication of critical regions, thus boosting both communication efficiency and decision-making efficacy. 
Empirical evaluations on the V2Xverse 
{and real-world DAIR-V2X 
datasets}
demonstrate that Select2Drive achieves a {$2.60$\% and $1.99$\%} 
improvement in offline perception tasks under limited bandwidth (resp., pose error conditions). 
Moreover, it delivers at most 
{$8.35$\% and $2.65$\%} 
enhancement in closed-loop driving scores and route completion rates, particularly in scenarios characterized by dense traffic and high-speed dynamics. 
\end{abstract}

\begin{IEEEkeywords}
Collaborative perception, Pragmatic communications, 
Data-based approaches, 
Connected and Autonomous Vehicles
%
\end{IEEEkeywords}

\section{Introduction}
Due to the inherent limitations of Autonomous Driving (AD), such as restricted visibility \cite{liu2023plant}, 
unpredictability of other road users \cite{Shao2023ReasonNet}, 
and difficulties in determining optimal paths \cite{peng2021learning}, 
Vehicle-to-Everything (V2X) Communications have become an indispensable ingredient in the Internet of Vehicles. By enabling the exchange of complementary information among vehicles, roadside units (RSUs), and even pedestrians, 
V2X communications promises a broadened perceptual horizon for individual autonomous vehicles \cite{Sedar2023V2X}, 
contributing to timely identification of emergent objects beyond visual observations \cite{xu2022opv2v} and swiftly making proper responses \cite{cui2022coopernaut}. 
Conventionally, early studies in the field of V2X communications focused on the realization of ubiquitous connectivity for accomplishing collaborative perception \cite{wang2020v2vnet}. 
However, the associated communication costs scale linearly with the size of the perceptual region and the time duration grows quadratically with the number of collaborating agents \cite{xu2022v2x}, 
placing significant demands on even next-generation communication systems \cite{3GPP_TS38.211}. 
Meanwhile, collaborative perception within a small number of neighboring agents and a limited timeframe only yields marginal performance improvement over single-agent perception \cite{hu2024pragmatic}. 
%
Fortunately, for V2X communications-assisted AD (V2X-AD), its sole reliance on reliable communications and the ignorance of the lasting impact of perception results on autonomous driving decisions still leave enormous room for optimization. 
\begin{figure}[tbp]
    \centering
    \includegraphics[width=1\linewidth]{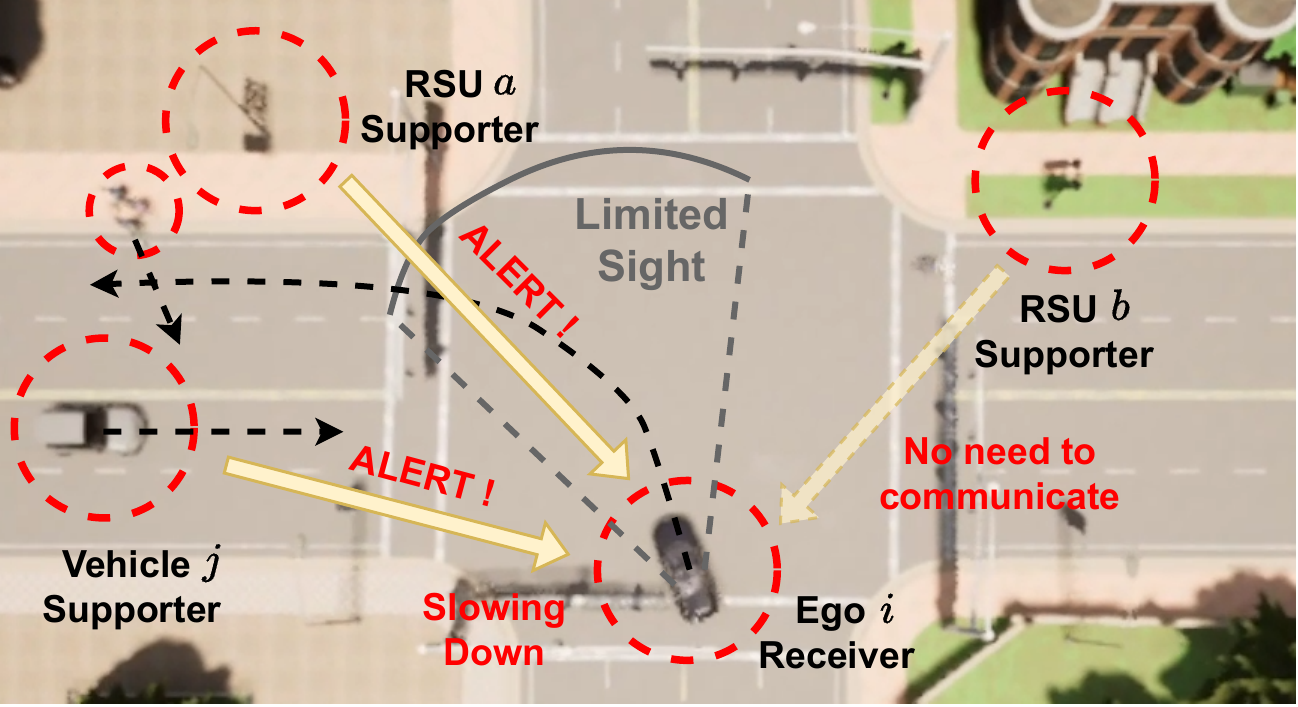}
    \caption{Overview of V2X-AD. Contingent on pragmatic  communications of driving-critical information with nearby supporters (e.g., vehicles and RSUs), the Ego vehicle maintains safe AD.} 
    \label{Fig.Overview}
    \vspace{-6mm}
\end{figure}

\begin{table*}[tbp]
    \centering
    \footnotesize
    %
    \caption{A comparison between Select2Drive and related works.}
    \begin{threeparttable}
    \begin{tabular}{|c|c|c|c|c|m{8cm}|}
        \toprule
        \textbf{References} & \begin{tabular}[c]{@{}c@{}}\textbf{Realistic} \\ \textbf{Communications}\end{tabular} & \begin{tabular}[c]{@{}c@{}}\textbf{Latency}\\ \textbf{Considered}\end{tabular} & \multicolumn{1}{l|}{\begin{tabular}[c]{@{}l@{}}\textbf{Perception}\\ \textbf{Involved}\end{tabular}} & \begin{tabular}[c]{@{}c@{}}\textbf{Driving-Task}\\ \textbf{Oriented}\end{tabular} & \textbf{Brief Description} \\ \midrule
            \textbf{\cite{liu2024towards}} & \Circle & \Circle & \CIRCLE & \CIRCLE & Integrates basic collaborative perception into closed-loop driving, lacks communication frameworks and real-world latency simulation. \\ \hline
            {\textbf{\cite{qu2024sicp}}} & \Circle & \Circle & \CIRCLE & \Circle & Proposes Dual-Perception Network (DP-Net), a lightweight network enabling simultaneous individual/cooperative 3D detection with State-Of-The-Art (SOTA) performance.
            \\ \hline
            \textbf{\cite{cui2022coopernaut}} & \CIRCLE & \Circle & \Circle & \CIRCLE & A blind-spot warning mechanism without engaging in precise collaborative perception and lack of generalization ability. \\ \hline
            \textbf{\cite{liu2020when2com,hu2022where2comm}} & \Circle & \Circle & \CIRCLE & \Circle & 
            Fetches the most valuable information for exchange under the premise of an ideal communication assumption, susceptible to latency issues. \\ \hline
            \textbf{\cite{Wei2023AsynchronyRobust,lei2024robust}} & \Circle & \CIRCLE & \CIRCLE & \Circle & A centralized estimation upon the timing of incoming information, imposes significant challenges on mobile devices' computational and storage capacities. \\ \hline
            \textbf{\cite{liu2024select2col}} & \CIRCLE & \CIRCLE & \CIRCLE & \Circle & A centralized, latency-based collaborator selection mechanism, incorporating the receiver's historical data into perception, proves inefficient in utilizing communication resources effectively. \\ \hline
            \textbf{Ours} & \CIRCLE & \CIRCLE & \CIRCLE & \CIRCLE & Implementation of a distributed prediction mechanism to mitigate overall latency, and pre-filtering invaluable information based on driving context before communication. \\ \bottomrule
        \end{tabular}
    \begin{tablenotes}
        \item \emph{Notations:} \Circle\ \emph{indicates not included;} \CIRCLE\ \emph{indicates fully included.}
    \end{tablenotes}
    \end{threeparttable}
    \label{tab.Innovation}
    \normalsize
    \vspace{-4mm}
\end{table*} 

Pragmatic Communications (PragComm), 
which aims to deliver compact latent representations tailored to specific downstream decision-making tasks, can better take into account both collaborative perception according to sensor data and subsequent driving decisions simultaneously\cite{10273599}.
Widely known as pragmatic compression or effective communications, the PragComm is commonly deployed as a compression paradigm in the context of V2X-AD \cite{hu2024pragmatic}. 
These methods operate under a fundamental assumption: during each time interval $\IntervalOverall$, all participating agents first broadcast Basic Safety Messages (BSMs) and subsequently decide whether to engage in communication \cite{liu2020when2com} or exchange valuable perception blocks \cite{hu2022where2comm}. 
However, this approach presumes an idealized scenario in which the entire process, regardless of the number of point-to-point communication links, can be completed within each $\IntervalOverall$. 
Apparently, this assumption is impractical due to inevitable latency from transmission and inference delays\footnote{Latency here specifically refers to the minimum response time required for a background vehicle to collect, process, and transmit data until the information is fused at the ego vehicle.}. 

On the other hand, despite advancements in collaborative perception, a critical gap lies in understanding how perception enhancements impact integrated, system-level driving performance. 
Typically, Imitation Learning (IL) \cite{liu2023plant} instead of Reinforcement Learning (RL) \cite{chen2020learning} is adopted owing to the remarkable performance of Behavior Cloning (BC) in accident scenarios on predefined routes.
Counterintuitively, as shown in findings from Ref. \cite{sun2025revisiting},  
particularly under augmented, collaborative perception, an expanded field of vision does not consistently improve decision-making, advocating for a ``\emph{less is more}" principle in V2X-AD. 
In other words, for closed-loop driving tasks, isolated perception modules often fail to seamlessly benefit subsequent planning and control stages, while incurring troublesome \emph{error propagation}, since inaccuracies in perception accumulate through the system \cite{10258330}. 
%
Therefore, in order to address latency-induced collaborative perception inconsistencies and ensure a consistent driving improvement, PragComm shall be redeveloped beyond simple context compression. 

In this paper, we propose Select2Drive, a revamped PragComm-based framework that not only accounts for the compensation of overall latency but also incorporates calibrations tailored for eliminating error propagation in V2X-AD.
Particularly, on top of a formulated delivery model that contributes to evaluating the underlying physical transmission plausibility \cite{10.1145}, Select2Drive introduces a novel Distributed Predictive Perception (DPP) module, which is capable of predicting future semantic features using low-level indicators. 
Notably, despite the conceptual simplicity, implementing DPP is non-trivial, as the limited computational capability requires precise forecasting of future states from high-dimensional voxel flow or pseudo-maps, focusing on minimizing disparities between predicted and current heatmaps. 
Furthermore, inspired by the underscored benefits of constrained observational horizons \cite{liu2023maskma}, 
Select2Drive investigates the feasibility of decision-making strategies using minimal observation content. 
This finally culminates in an Area-of-Importance-based PragComm (APC) framework, which prioritizes communications in driving-critical regions. 
While providing key distinctions with highly relevant literature in Table \ref{tab.Innovation}, our key contribution could be summarized as follows: 
\begin{itemize}  
\item To significantly boost the closed-loop driving performance under the impact of communication and computational latency, we propose a PragComm-based, IL-enabled real-time collaborative driving framework Select2Drive. Beyond simple information compression, the DPP and APC components therein can effectively incorporate background vehicle information while avoiding redundant computational burden and minimizing unnecessary communication.
\item The calibrated DPP component integrates a predictive mechanism and a motion-aware affine transformation, which leverages low-dimensional motion flow to infer future semantic features. Avoiding direct prediction of high-dimensional Bird's Eye View (BEV) semantic features effectively mitigates timeliness challenges without introducing significant computational cost.
\item Bearing the ``\emph{less is more}'' principle in mind, we introduce a revamped APC component that restricts the communication region to the Area-of-Importance (AoIm), effectively alleviating the \emph{covariate shift} induced by BC on constrained datasets. 
Therefore, Select2Drive enables prioritized communication in driving-critical regions and solves the latency-induced fusion inconsistencies from collaborative perception. 
\item Building upon the CARLA Simulator \cite{dosovitskiy2017carla} and prior studies \cite{liu2024towards}, we develop a comprehensive simulation platform\footnote{The open-source codes can be found at \url{https://github.com/zjunice} once the manuscript has been accepted.} 
that transitions collaborative perception approaches from offline datasets to closed-loop driving scenarios \cite{carlaleaderboard} while offering an extensible interface for multi-vehicle cooperative driving. 
Through extensive experiments on both collaborative perception tasks and online closed-loop driving tasks, we demonstrate significantly improved performance (e.g., {$2.60\%$} higher perception accuracy in a simulated dataset V2Xverse, {$1.99\%$ higher perception accuracy in a real-world dataset DAIR-V2X}, {$8.35\%$} higher closed-loop driving scores, and {$2.65\%$} larger route completion rates) of Select2Drive across diverse communications-limited scenarios. 
\end{itemize}  

The remainder of this paper is organized as follows.
We introduce related works in Section \ref{sec2}.
We introduce system models and formulate the problem in Section \ref{sec3}.
In Section \ref{sec4}, we elaborate on the details of our proposed prediction paradigm.
In Section \ref{sec5}, we present the experimental results and discussions.
Finally, Section \ref{sec6} concludes this paper. 
\begin{figure*}[tbp]
    \centering
    \includegraphics[width=\linewidth]{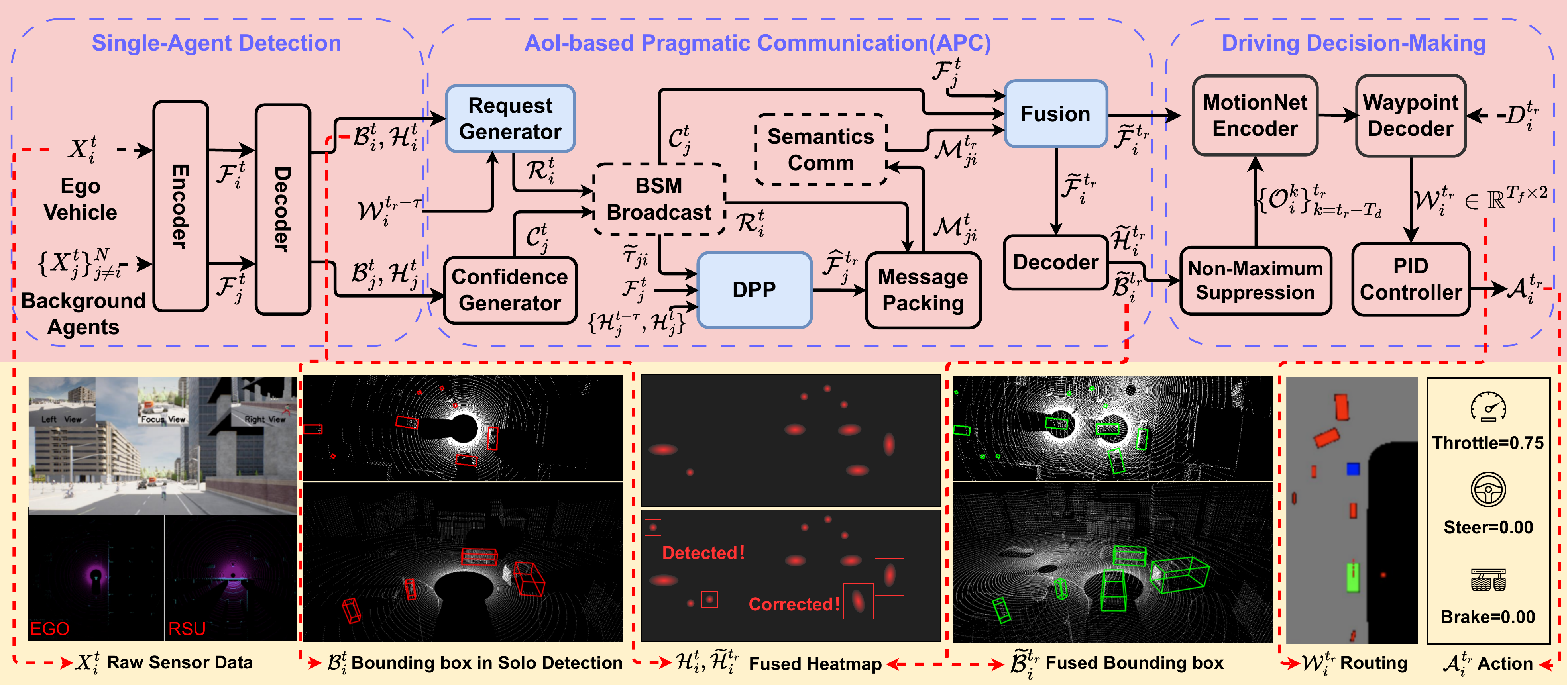}
    \caption{System model of our V2X-AD framework encompassing perception, decision-making, and control stages. 
    The upper section provides a detailed closed-loop flowchart, illustrating the iterative cycle from perception to action, while incorporating feedback into subsequent iterations. 
    The lower section visually depicts the complete decision-making process, emphasizing the sequential flow of actions and data exchange.} 
    \label{Fig.SystemModel}
    \vspace{-4mm}
\end{figure*}

\section{Related Works}\label{sec2}
\subsection{End-to-end Autonomous Driving}
Recent advancement in learning-based end-to-end autonomous driving, which directly translates environmental observations into control signals \cite{liu2023plant} and conceptually addresses the cascading errors of traditional modular design \cite{shao2022interfuser}, has positioned this domain as a pivotal research focus. 
%
Nevertheless, existing methods highlight a gap between theoretical assumptions and practical implementation. For example, Ref. \cite{hao2025research} demonstrates the performance of collaborative perception algorithms in simulated environments but these are rarely applied to real-world driving tasks. 
Ref. \cite{10229239} assumes accurate agent position data, which is often impractical in real-world scenarios. 
%
Our approach bridges this gap by integrating theoretical strategies with higher-fidelity implementations, which utilize perception data directly from emulated raw sensor inputs for more realistic analysis.

\textbf{Learning Approaches:} 
End-to-end driving approaches can be classified into RL-based or supervised learning-based IL \cite{10258330} \footnote{Traditional RL/IL methods are typically limited to lower-dimensional problems. Therefore, the methodologies discussed in this paper specifically refer to Deep Learning (DL)-driven RL/IL approaches.}. 
Compared to RL-based solutions, IL progressively benefits from the increasing perception performance, leading to a stable enhancement in the learning of driving tasks through BC \cite{liu2023plant}.
Notably, BC demonstrates effective performance for in-distribution states within the training dataset but struggles to generalize to Out-Of-Distribution (OOD) states due to compounding action errors, a phenomenon termed \emph{covariate shift} \cite{ross2011reduction}. 
%
To mitigate this, we intentionally add noise to expert control signal to ensure more states within the training distribution \cite{8460487}. 

\textbf{AD:} 
Ref. \cite{cui2022coopernaut} proposes a visually cooperative driving framework that aggregates voxel representations from multiple collaborators to improve decision-making. 
Ref. \cite{Shao2023ReasonNet} demonstrates that besides challenges in predicting the motion of out-of-view or non-interactive objects, single-agent driving systems inherently struggle with occluded or distant regions, often leading to catastrophic failures.
To address these limitations, V2X-AD adopts a multi-agent collaborative paradigm leveraging V2X communications, enabling vehicles to share information and collaboratively make informed decisions \cite{hao2025research}. Despite the remarkable progress, the latest evaluation platform \cite{liu2024towards} remains constrained by idealized communication assumptions.

\subsection{Pragmatic Communications}
\label{sec:prag_comm}
Commonly formulated as an extension of the Markov Decision Process (MDP) framework \cite{9466501}, PragComm shifts the focus from accurate bit transmission or precise semantic interpretation to capturing key information and creating compact representations for specific downstream tasks. 

\textbf{V2X Communications:} 
{Information exchange for V2X cooperation can be posed as an image-transmission task whereby vehicles periodically capture and disseminate camera frames. Considering RGB image sharing, a front-view camera operating at $10$ Hz with $2048\times1024$ resolution and $24$-bit color depth produces approximately $48$ Mb per frame; lossless PNG compression reduces this to about $18.85$ Mb \cite{gimenez2024semantic}.
%
As of 2024, 3GPP specifies up to $53$ Mbps for User Equipment (UEs) information sharing in V2X applications \cite{3GPP_TS38.211}, implying a maximum image-sharing rate of roughly $2.81$ Hz, which is insufficient for exchanging raw camera data in the near term. 
Therefore, efficient filtering and compression of perception data are essential for real-time performance.}
%
PragComm is contingent on the underlying capability of V2X communications, such as IEEE 802.11p-based DSRC \cite{IEEE802} and the 3GPP Cellular-based V2X (C-V2X) \cite{7992934}. 
Both architectures define BSMs \cite{kenney2011dedicated, 9212349}, transmitted periodically at up to 10 Hz to convey critical state information such as position, dynamics, and vehicle status. 
Correspondingly, high-frequency BSMs can serve as a foundation for high-dimensional semantic feature communication, minimizing redundant transmissions. For DSRC-based transmission, bandwidth-limited channel conditions highlight the necessity to investigate the impact of communication latency on collaborative perception, while the reliance on inter-node routing in C-V2X-based transmission necessitates a focus on systemic overall delays. 

\textbf{PragComm in V2X-AD:} 
Ref. \cite{hu2022where2comm} establishes a PragComm-based framework towards achieving a balance between perception performance and communication costs in V2X-AD. 
It employs a two-step strategy: (1) semantic feature extraction from raw sensory data to low-level heatmaps as indicators; (2) selective transmission of high-value semantic features for fusion to optimize communication efficiency.  
However, considering the heterogeneity in distance and content, PragComm in V2X-AD encounters difficulties spanning from localization uncertainty \cite{lei2024robust} to clock synchronization and dynamic delay compensation \cite{Wei2023AsynchronyRobust}. 
For example, even minimal delays can profoundly undermine the timeliness of transmitted information, potentially incurring catastrophic outcomes \cite{lei2022latency}. 
%
Meanwhile, prior methodologies primarily focus on reconstructing the distribution of proximal objects. 
While enhancing perception, these methods often misalign with driving policy optimization, necessitating integrated frameworks for cohesive performance.
In that regard, Ref. \cite{9466501} underscores that the decoupling of learning and communication yields suboptimal results. 
Therefore, there emerges a strong incentive to revamp PragComm for AD.  

Compared to the literature, Select2Drive employs DPP, which diverges from traditional approaches by integrating a prediction mechanism at the supporter level to alleviate the impact of inevitable delays without imposing considerable computational burdens. 
Meanwhile, Select2Drive takes advantage of APC to bridge the disconnection between perception modules and low-level controllers by explicitly incorporating prior trajectory information into communication strategies. Therefore, Select2Drive not only further minimizes communication overhead but also sharpens the model's focus on task-critical information, ultimately enhancing driving performance.

\begin{table}[tbp]
    
    \centering
    \caption{A summary of major notations used in this paper.}
    \vspace{-0.5em}
    \begin{tabular}{|c|m{6cm}|}
        \toprule
        \textbf{Notation} & \textbf{Definition} \\ \midrule
        $\SensorEgo, \PseuEgo$ & Raw sensor data and latest available semantic features of agent $i$ at time $t$ \\ \hline
        $\HeatMapEgo, \BoundingBoxEgo$ & Heatmap and bounding box from agent $i$\\ \hline
        $\HeatMapBackHistoryBatchAll$ & Historical heatmaps from agent $j$\\ \hline
        $\HeatMapProcedure{\ReceiveTime}, \PseuBackProcessed$ & Forecasted heatmap and processed semantic features from agent $j$\\ \hline
        $\ConfidenceMapBack, \RequestMapEgo$ & Confidence map from agent $j$ and request map from agent $i$ \\ \hline
        $\PseuFused, \HeatMapFused, \BoundingBoxFused$ & Fused semantic feature and collaborated perception of agent $i$\\ \hline
        $\OccupancyMapEgoHistory$ & $\LenDrivingMemory$ frames of historical Bird's Eye View (BEV) occupancy maps in the view of agent $i$\\ \hline
        $\Waypoints, \DriveAction$ & Estimated trajectory and expected driving action of agent $i$\\ \hline
        $\IntervalSync$ & Broadcast period of request map $\RequestMapEgo$\\ \hline
        $\LatencyOther, \LatencyOtherEst$ & Overall transmission latency between agent $j$ and agent $i$, and the related estimation \\ \hline
        $\LatencyOverall, \LatencyOverallEst$ & Real systematic latency between agent $j$ and agent $i$, and the related estimation \\ \bottomrule
    \end{tabular}
    \vspace{-4mm}
    \label{tab.notions}
\end{table}

\section{System Model and Problem Formulation}\label{sec3}
%
Beforehand, primary notations used in this paper are summarized in Table \ref{tab.notions}. 
In the subsequent discourse, intermediate variables output by the Deep Neural Network (DNN) will be denoted using a script font (e.g., $\PseuEgo$), while directly observable variables will be represented in a standard font (e.g., $\Navigation$). In addition, a DNN will be denoted as a function $\Phi(\cdot)$.

\subsection{System Model}
In this paper, we consider a collaborative perception-based AD scenario with multiple vehicles (i.e., agents). Particularly, as shown in Fig. \ref{Fig.SystemModel}, let $t$ represent the moment when an agent $i$ initiates a decision-making cycle. At time $t$,  agent $i$ can perceive raw data (e.g., RGB images and 3D point clouds) at a fixed interval $\IntervalOverall$, while the communications possibly occur between any ego agent $i$ and one of its supporting neighbors $j$ (i.e., background vehicles and RSUs). Afterwards, agent $i$ aims to maximize the accomplishment rate of its IL-based driving task with driving plan $\Waypoints$, contingent on the fusion of its own observed raw data $\SensorEgo$ and exchanged information $\{\Msg\}_{\NAgents}$ from neighboring agents $j \in \NAgents$. Basically, such a scenario can be classified as a pragmatic communications-based MDP. 
\subsubsection{Confidence-Driven Message Packing}
After obtaining the raw sensor data $\SensorEgo$, each vehicle leverages an encoder $\ModEncoder$, which consists of a series of 2D convolutions and max-pooling layers, to yield the latest available semantic features $\PseuEgo$ that merge RGB images and 3D point clouds into a unified global coordinate system, namely 
\begin{equation}  
    \PseuEgo = \ModEncoder(\SensorEgo) \in \mathbb{R}^{\Height \times \Width \times \Channel},  
    \label{Eq.extraction}  
\end{equation}  
where $\Height$, $\Width$, and $\Channel$ denote the dimensions of the pseudo-image. Typically, $\Channel \gg 1$ even only 3D point clouds are utilized. 
Subsequently, a decoder $\ModDecoder$, composed of several deconvolution layers, is employed to generate a probability heatmap $\HeatMapEgo$ and a bounding box regression map $\BoundingBoxEgo$. The heatmap $\HeatMapEgo$ represents the spatial likelihood of an object (e.g., vehicles, pedestrians, or traffic signs) being present in an image or frame, while the regression map $\BoundingBoxEgo$ provides precise localization details  (e.g., center coordinates, width, and height) for detected objects. Therefore, 
\begin{equation}  
    \HeatMapEgo, \BoundingBoxEgo  = \ModDecoder(\PseuEgo) \in \mathbb{R}^{\Height \times \Width \times \NumClass}, \mathbb{R}^{\Height \times \Width \times 8\NumClass},  
    \label{Eq.decoder}  
\end{equation}  
with $\NumClass$ representing the number of object categories and $\NumClass = 3$ if three categories, i.e. vehicles, bicycles, and pedestrians, are detected.

Afterward, the agent $i$ sends low-dimensional BSMs, including an $\ConfidenceMapGenerator$-induced confidence map $\ConfidenceMapEgo$ and a request map $\RequestMapEgo$, as:
\begin{align}
    & \ConfidenceMapEgo = \ConfidenceMapGenerator(\HeatMapEgo) \in [0, 1]^{\Height \times \Width},  \\ 
    & \RequestMapEgo = 1 - \ConfidenceMapEgo \in [0, 1]^{\Height \times \Width}, 
    \label{Eq.confidence}
\end{align}  
where $\ConfidenceMapGenerator$ denotes a maximum operation in the third dimension followed by a Gaussian filter. 
Under the ideal latency-free assumption, for the supporting vehicle $j \in \NAgents$, confidence-driven messages for feedback are given as: 
\begin{equation}
    \Msg = \PseuBack \times \PassingMask \in \mathbb{R}^{\Height \times \Width\times\Channel}. 
    \label{Eq.MessagePacking}
\end{equation} 
Here, $\PassingMask = \OneMask{\RequestMapEgo \odot \ConfidenceMapBack \geq \ThresholdCollab } \in \mathbb{R}^{\Height \times \Width}$ indicates a spatial selection mechanism for $\PseuBack$, 
and $\ThresholdCollab$ is a hyperparameter controlling the extent of collaboration. The indicator $\OneMask{\cdot}$ equals $1$ if the condition is met; while nulls otherwise. The operator $\odot$ denotes element-wise multiplication. 

\begin{figure}[tbp]
    \centering
    \includegraphics[width=1\linewidth]{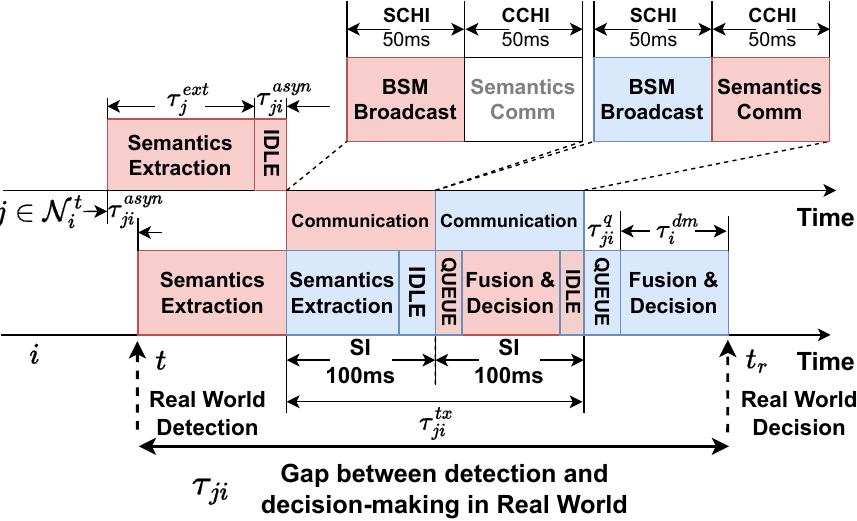}
    \caption{{Flow chart from the perspective of agent $i$.} The \textcolor{red}{red} box delineates the decision cycle initiated at time $t$, while the \textcolor{blue}{blue} box represents the subsequent cycle commencing at time $t+\IntervalOverall$. The interval between consecutive perception and communication phases is uniformly set to $\IntervalOverall$. }
    \label{Fig.delivery}
    \vspace{-4mm}
\end{figure} 
\subsubsection{Latency Model} \label{sec.latency}
The acquisition, communication and post-processing of $\Msg$ inevitably incur some latency, such as the computational latency involved in semantic extraction $\LatencyExt$ and post-processing for decision-making $\LatencyDecision$, the asynchronous inter-agent timing differences {and latency jitter} $\LatencyAsync$, and the more prominent communication latency\footnote{Notably, due to the low-dimensional nature of BSMs, the latency for transmitting BSMs is assumed to be negligible relative to that for transmitting high-dimensional semantics $\Msg$ (i.e., $\LatencyOverall$).} $\LatencyTrans$.

As depicted in Fig. \ref{Fig.delivery}, to quantify $\LatencyTrans$, a verification mechanism proposed in \cite{8356112} is employed. Notably, in the multi-channel alternating switch mode therein, the communication process is structured into a Synchronization Interval (SI), denoted as $\IntervalOverall$, which is further divided into a Service Channel Interval (SCHI) and a Control Channel Interval (CCHI), each lasting $\IntervalSync$. During the SCHI, BSMs, such as $(\RequestMapEgo$ and $ \ConfidenceMapEgo)$ are broadcast, while semantic information $\Msg$ is transmitted during the subsequent CCHI. 
The minimum transmission time for $\Msg$ is given by  
\begin{equation}
    \LatencyTrans = \LatencyPropagation + \LatencyNetwork. 
\end{equation} 
Here, $\LatencyPropagation$ represents the propagation latency, computed as per 3GPP TR 38.901 \cite{3GPP}, that is, 
\begin{equation}  
\LatencyPropagation = \text{size}({\Msg})\bigg/ \left(\BandWidthSingle \log_2(1+10^{0.1(\TransPower-\PathLoss-\TransNoise)})\right),
\end{equation}  
where $\BandWidthSingle$ is the bandwidth allocated per agent, $\TransPower$ is the transmission power, $\TransNoise$ is the noise power, and $\PathLoss$ denotes the path loss calculated by $\PathLoss = 28 + 22\log_{10}(\Distance) + 20\log_{10}(\CarrierFrequency)$ \cite{3GPP}, with $\Distance$ being the inter-agent distance (in meters) and $\CarrierFrequency$ the carrier frequency (in GHz).  
The term $\LatencyNetwork$ accounts for the processing time at network nodes (e.g., routers, switches, base stations) before forwarding data to the next hop. 
Owing to the direct communication characteristics of DSRC-based transmission, $\LatencyTrans$ is predominantly constrained by $\LatencyPropagation$, with negligible $\LatencyNetwork$ \cite{Sedar2023V2X}. 
Conversely, the multicast service in C-V2X-based transmission effectively mitigates $\LatencyPropagation$ while introducing substantial $\LatencyNetwork$ delays, primarily attributed to computational burdens at network nodes caused by access and handover overhead \cite{Gya2021V2X}.

In a nutshell, the overall latency $\LatencyOverall$ can be expressed as:  
\begin{equation}  
    \LatencyOverall = \LatencyExt  + \LatencyAsync + \LatencyTrans + \LatencyDecision {+ \LatencyQueue},
    \label{Eq.latency}  
\end{equation} 
%
{where the term $\LatencyQueue$ denotes the queueing latency \cite{gimenez2024semantic} for the ego agent to sequentially process multiple agent interactions.}
For notational simplicity, $\ReceiveTime$ denotes the moment when the agent $i$ obtains the message $\Msg$ from another agent $j$ during the cycle starting at $t$. Thus, $\ReceiveTime = t + \LatencyOverall$. 

%
\subsubsection{Information Fusion and Decision-Making}
\label{sec:fusion}
After the communications, the ego vehicle would aggregate all available information $\MsgAll$\footnote{
For simplicity of representation, we denote $\MsgSelf = \PseuEgo$ and $\LatencyOverallSelf = 0$.} to derive fused features $\PseuFused$ 
\begin{equation}
    \PseuFused = \ModFuse\left( \{\WeightFusion \odot \Msg\}_{\AllAgents} \right)\in \mathbb{R}^{\Height \times \Width \times \Channel}, 
    \label{Eq.fusion_MHA}
\end{equation}
where $\ModFuse$ is implemented with a feed-forward network and $\WeightFusion = \ModMHA \left( \PseuEgo, \Msg, \Msg \right) \odot \ConfidenceMapBack \in \mathbb{R}^{\Height \times \Width}$ indicates a Scale-Dot Product Attention (SDPA)\cite{Ashish2017Attention} generated with per-location multi-head attention  $\ModMHA$.
Next, by decoding the fused feature $\PseuFused$ through a predefined decoder $\ModDecoder$ as: 
\begin{equation}
    \HeatMapFused, \BoundingBoxFused = \ModDecoder(\PseuFused) \in \mathbb{R}^{\Height \times \Width \times \NumClass}, \mathbb{R}^{\Height \times \Width \times 8\NumClass}, 
    \label{Eq.decode_fused}
\end{equation}
where $\HeatMapFused$ and $\BoundingBoxFused$ represent the heatmap and bounding box regression map obtained with fused semantic information $\PseuFused$, respectively. 
3D objects are then detected via non-maximum suppression \cite{NMS} and rasterized into a binary BEV occupancy map $\OccupancyMapEgo$. Using $\LenDrivingMemory$-length historical occupancy maps $\OccupancyMapEgoHistory$ as well as the navigation information $\Navigation$, the ego vehicle leverages a learnable planner $\ModPlan$ encompassing a MotionNet encoder, a goal encoder and corresponding waypoint decoder \cite{shao2022interfuser} to generate a driving plan consisting of a series of waypoints $\Waypoints$. Mathematically, it can be described as:
\begin{equation}
    \Waypoints = \ModPlan(\{\HeatMapFused, \BoundingBoxFused\}_{k=1}^{\LenDrivingMemory}, \PseuFused, \Navigation) \in \mathbb{R}^{2\times\LenDrivingTrajectory}. 
    \label{Eq.plan}
\end{equation} 
The optimal driving action $\DriveAction$, comprising steering, throttle, and brake commands, is then determined via lateral and longitudinal Proportional–Integral–Derivative (PID) controllers $\ModController$ as:
\begin{equation}
    \DriveAction = \ModController(\Waypoints) \in [0,1]^{2} \cup \{0,1\}^1. 
    \label{Eq.control}
\end{equation} 
\subsection{Problem Formulation}
%
\begin{figure*}[htbp]
	\centering
    \subfloat[Main architecture]{\includegraphics[scale=0.5]{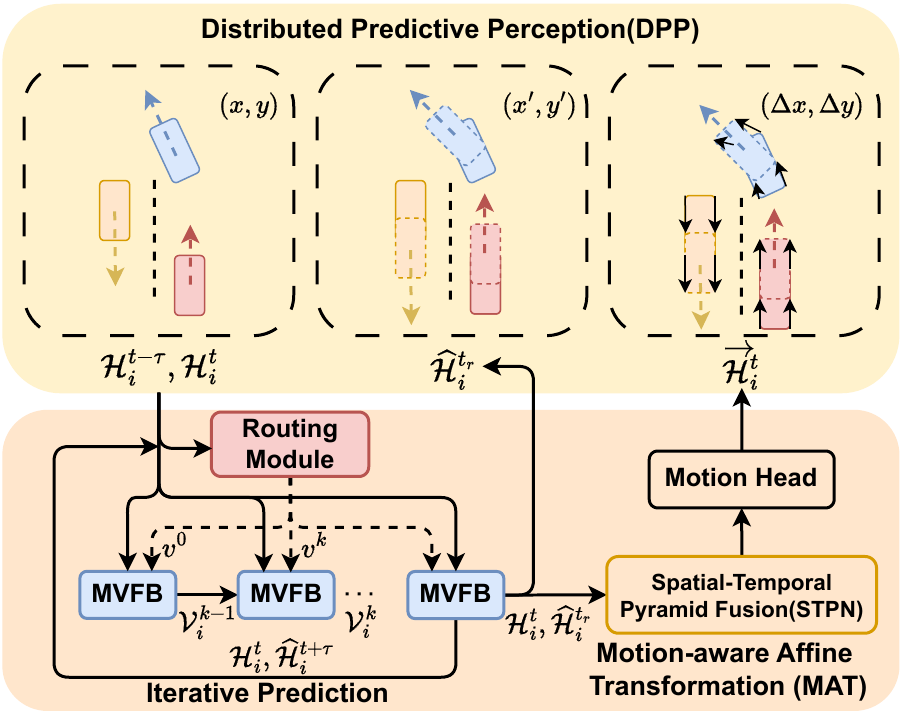}}
    \subfloat[$k$-th MVFB]{\includegraphics[scale=0.5]{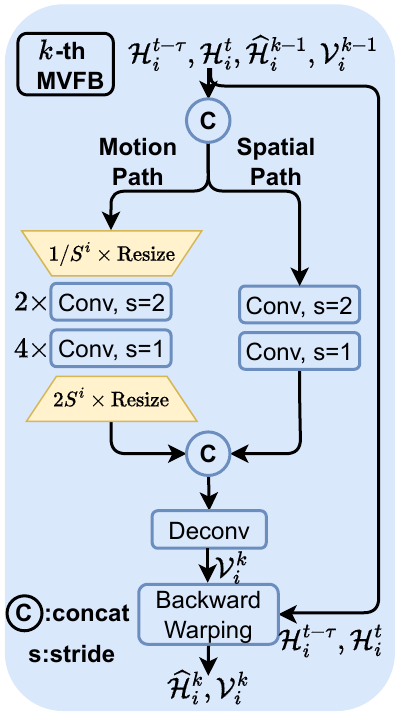}}
    \subfloat[Routing Module]{\includegraphics[scale=0.5]{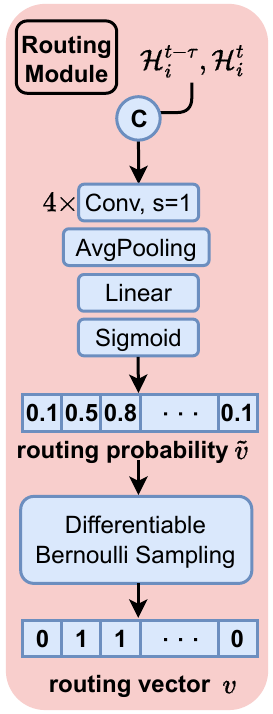}}
    \subfloat[STPN and Motion Head]{\includegraphics[scale=0.5]{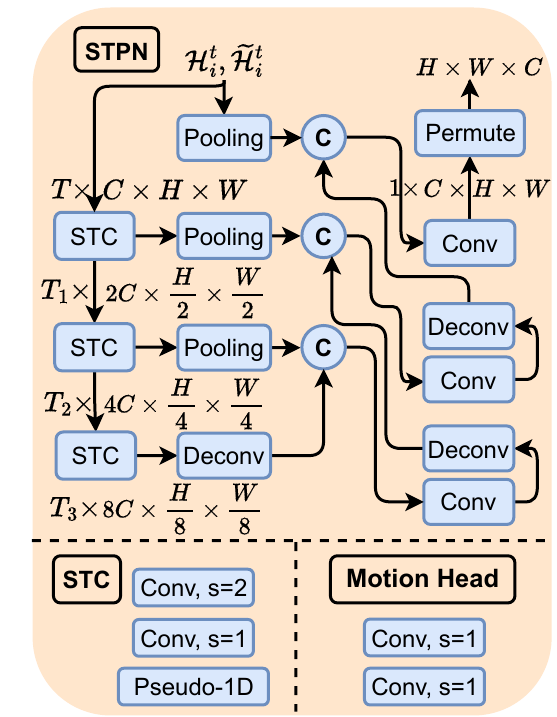}}
    \caption{Overview of the proposed DPP framework. }
    \label{Fig.DPP}
\end{figure*}

This paper aims to maximize the achievable driving performance through calibrated pragmatic communication. Particularly, the PragComm-based V2X-AD problem can be consistently formulated as:  
\begin{align}
    \max_{\ParamPlan, \ParamPerception} & \sum_{i=1}^N \EvalDrive \left[ \Waypoints , \ModPlan \left( \ModPerception(\SensorEgo, \{\Msg\}_{\NAgents}) \right) \right],\nonumber\\
    \text{s.t.}\ &\Msg = \ModProcess(\PseuBack, \HeatMapEgo), \label{Eq.objective} \\
    & \abs{\Msg}\leq \BandWidthSingle*\IntervalSync\ \text{for}\ j\in\NAgents,\nonumber
\end{align}  
where $\ModPerception$ represents all involved perception-related DNNs in Eqs. \eqref{Eq.extraction} to \eqref{Eq.decode_fused}. 
Specially, $\ModProcess$ corresponds to the DNN-based PragComm components outlined in Eqs. \eqref{Eq.confidence} and \eqref{Eq.MessagePacking}, and the operator $\EvalDrive(\cdot)$ indicates metrics \cite{jia2024bench2drive} (e.g., route completion rates, infraction penalty and driving scores) in the driving scenarios, considering both safety rate and traffic efficiency. 
While the request map, as derived in Eq. \eqref{Eq.confidence}, provides an intuitive foundation, it lacks task-specific optimization, such as the prioritization of information relevant to navigation information $\Navigation$, route planning \cite{wu2022tcp}, or salient objects \cite{SCOUT}.
%
Even worse, as mentioned earlier in Section \ref{sec:prag_comm}, under ideal channel conditions, communication resources are insufficient to achieve latency-free communication, even for extremely compressed messages. 

On the other hand, for the highly volatile AD environment, due to the existence of computation and communications latency $\LatencyOverall$, the currently available observations in Eq. \eqref{Eq.MessagePacking} might become outdated at time $\ReceiveTime$. 
Instead, directly transmitting the forecast semantic features, predicted at time $t$, to complement the possible impact of latency $\LatencyOverall$ is preferable. 
Nevertheless, although the estimation of the overall latency $\LatencyOverallEst$ is achievable through a synchronization mechanism as in Section~\ref{sec.latency}, the prediction of high-dimensional $\PseuBack$ might impose a significant computational burden on the mobile device. Therefore, beyond simple information compression, $\ModProcess$ (particularly Eq. \eqref{Eq.MessagePacking}) shall be carefully investigated to effectively incorporate predicted background vehicle information at the expense of reduced computational overhead and minimal communications. 

\section{Select2Drive: driving-oriented collaborative perception}\label{sec4}
In this section, we introduce a Select2Drive  framework that prioritizes the communication of decision-critical, timely content into the collaborative driving process. 
To obtain computationally efficient prediction, we reformulate it as a dimensionality reduction-based reconstruction problem and devise a DPP to extract the inherent transformation $\HeatMapFlowGT$, which represents the motion flow of objects from $\PseuBack$ to $\PseuBackToPredict$, and subsequently infer $\PseuBackToPredict$ from an affine approximation of $\PseuBack$. 
Furthermore, to ensure that improvements in perception performance  consistently translate to enhanced outcomes in offline driving simulations, 
we design the message-packing mechanism (i.e., APC) on top of DPP.

\subsection{Distributed Predictive Perception (DPP)}
As illustrated in Fig. \ref{Fig.DPP}, instead of directly predicting $\HeatMapFlowGT$ from high-level semantics $\PseuBack$, which exhibits significant sensitivity to continuous latency $\LatencyOverall$, 
we first downsample the semantics $\{\PseuBack, \PseuBackToPredict\}$ to low-level heatmap \cite{NEURIPS2023_6ca5d266} $\{\HeatMapBack, \HeatMapBackToPredict\}$ with the decoder $\ModDecoder$ in Eq. \eqref{Eq.decoder}. As mentioned earlier, due to the temporary unavailability of $\HeatMapBackToPredict$, we leverage a video prediction-inspired \emph{iterative prediction} method to learn a predicted version $\HeatMapProcedure{\ReceiveTime}$. 
Next, we introduce a \emph{motion-aware affine transformation} mechanism to extract motion information $\HeatMapFlow \in \mathbb{R}^{\Height \times \Width \times 2}$, which corresponds to the $2$-dimensional positional shifts $(\Delta x, \Delta y)$ for every object initially located at $(x, y)$.
%
%
%
\begin{table*}[tbp]
    \centering
    \caption{Parameters and computational overhead of major modules. {Modules highlighted with \textbf{bold} are selected as the backbone of our model for integration to meet the $100$ ms decision interval requirement.}}
    \begin{threeparttable}
    \begin{tabular}{|c|ccccc|}
    \hline
    \textbf{Module} & {\textbf{Params} (M)} & {\textbf{FLOPs} (G)} & {\begin{tabular}[c]{@{}l@{}}\textbf{Execution time}\\ \textbf{on Desktop} (ms)\end{tabular}} & {\begin{tabular}[c]{@{}l@{}}{\textbf{Execution time}} \\ {\textbf{on Vehicle} (ms)}\end{tabular}} & {$\Delta$ \textbf{Performance to Ours} (\%)} \\ \hline
    \multicolumn{6}{|c|}{\textbf{Future Confidence Forecast Module ($\ModForecast$)}} \\ \hline
    {\textbf{DMVFN} \cite{2023DMVFN}} & {$3.6$} & {$2.1$} & {$1.17$} & {$1.10$} & $0$  \\ \hline
    {PredRNN++ \cite{2017PredRNN}} & {$24.6$} & {$169.8$} & {$94.80$} & {$89.45$} & $+0.62$  \\ \hline
    {TAU \cite{2023TAU}} & {$38.7$} & {$85.0$} & {$47.45$} & {$44.77$} & $+0.77$  \\ \hline
    {MAU \cite{2021MAU}} & {$10.5$} & {$29.1$} & {$16.25$} & {$15.33$} & $+0.27$  \\ \hline
    {PhyDNet \cite{2020PhyDNet}} & {$5.8$} & {$80.7$} & {$45.00$} & {$42.46$} & $+0.44$  \\ \hline
    \multicolumn{6}{|c|}{\textbf{Semantic Feature Extraction Module ($\ModEncoder$, $\ModDecoder$)}} \\ \hline
    {\textbf{PointPillar} \cite{lang2019pointpillars}} & {$8.2$} & {$119$} & {$66.46$} & {$62.71$} & $0$  \\ \hline
    {{CenterPoint \cite{yin2021center}}} & {$8.2$} & {$170$} & {$94.94$} & {$89.58$} & $+0.10$  \\ \hline 
    \multicolumn{6}{|c|}{\textbf{Motion Perception Module ($\ModMAT$, $\ModPlan$)}} \\ \hline
    {\textbf{MotionNet} \cite{wu2020motionnet}} & {$1.7$} & {$10.2$} & {$5.70$} & {$5.38$} & $0$  \\ \hline
    {{LSTM}} & {$3.5$} & {$30.46$} & {$17.02$} & {$16.06$} & $-4.56$  \\ \hline
    \multicolumn{6}{|c|}{\textbf{Intermediate Feature Fusion Module ($\ModFuse$)}} \\ \hline
    {\textbf{SDPA} \cite{Ashish2017Attention}} & {$0.007$} & {$0.29$} & {$0.16$} & {$0.15$} & $0$  \\ \hline
    {Max Fusion} & {$0$} & {$0$}& {$0.05$} & {$0.05$} & $-3.07$  \\ \hline
    \end{tabular}
    \begin{tablenotes}
        \footnotesize 
        \item[1] {Execution time on Desktop is measured on an RTX 4090 ($1,321$ TOPS) in single-step, one-to-one driving scenarios. Vehicle time is estimated by scaling against NVIDIA THOR [57] ($2,000$ TOPS) with $70\%$ utilization to account for operating system overhead. Notably, Max Fusion solely involves element-wise maximum operations without floating-point computations, resulting in null FLOPs.} 
        \item[2] {In the context of the perception task, ``$\Delta$ Performance to Ours" quantifies the performance gap when a specific module is substituted, with our method serving as the established baseline.} 
        \normalsize
    \end{tablenotes}
    \end{threeparttable}
    \label{tab.modules}
    \vspace{-4mm}
\end{table*}
%
%

\subsubsection{Iterative Prediction} 
We discretize the estimated latency $\LatencyOverallEst$ into discrete steps $\LatencySteps = \lfloor \LatencyOverallEst / \IntervalOverall \rfloor$ \cite{lei2022latency}. 
Thus, $\ReceiveTimeEstimate = t + \LatencySteps \times \IntervalOverall$, consistent with the decision-making cycle $\IntervalOverall$ in Section \ref{sec.latency}. 
We iteratively generate a heatmap sequence $\{\HeatMapProcedure{t+\IntervalOverall}, \dots, \HeatMapProcedure{\ReceiveTimeEstimate}\}$ through $\LatencySteps$ steps as: 
\begin{equation}
    \HeatMapProcedure{t+\IntervalOverall}, \dots, \HeatMapProcedure{\ReceiveTimeEstimate} \ce{<=>[iteratively][for \mathit{\LatencySteps}]} \ModForecast(\HeatMapBackHistoryBatchAll),  
    \label{Eq.DPP}
\end{equation}
\vspace{-.1cm}

\noindent while employing $\HeatMapProcedure{\ReceiveTimeEstimate}$ to approximate $\HeatMapProcedure{\ReceiveTime}$.

{Before delving into the implementation details, Table \ref{tab.modules} summarizes popular candidates and compares model parameter count, computational complexity (measured in FLOPs per inference), mean latency (evaluated empirically and estimated on the vehicle platform \cite{NVIDIATHOR}), and the performance impact of individual module modifications.
In short, we first prioritize real-world deployability under a $10$ Hz decision frequency, and subsequently select the optimal models in the end-to-end perception task.
Such a procedure leads to an integration of PointPillar \cite{lang2019pointpillars}, DMVFN \cite{2023DMVFN}, MotionNet \cite{wu2020motionnet}, and SDPA \cite{Ashish2017Attention}.
Notably, it trades a maximum $0.87\%$ 
performance loss for a $50.43\%$ (resp., $70.54$ ms) 
reduction in decision-making latency on vehicles, resulting in a total latency of $69.34$ ms.} 
Specifically, DMVFN excels by operating without extra inputs and avoiding redundant convolutions, making it ideal for dense decision-making in autonomous driving.

As depicted in Fig. \ref{Fig.DPP}, to ensure remarkable performance in resource-constrained settings, the DMVFN employs $K = 9$ Multi-scale Voxel Flow Blocks (MVFBs) coupled with a dynamic routing module. 
Particularly, to effectively capture large-scale motion while maintaining spatial fidelity, each MVFB $k \in \{1,\cdots, K\}$ incorporates a dual-branch network structure, encompassing a motion path and a spatial path, to downsample the inputs by a scaling factor $\ScalingFactor$ for a larger receptive field while preserving fine-grained spatial details \cite{8237740}. 
Subsequently, the outputs from both paths are concatenated to predict the voxel flow $\VoxelFlow{k}$, which is then applied through backward warping \cite{NIPS2015_33ceb07b} to generate a synthesized frame $\HeatMapProcedure{k}$. 

Without loss of generality, taking the example of inputting $(\HeatMapBackHistoryBatchAll)$, each MVFB $k$ is achieved by processing these two historical frames, a synthesized frame $\HeatMapProcedure{k-1}$ and the voxel flow $\VoxelFlow{k-1}$ generated by the $k-1$ MVFB block. Thus, we have
\begin{equation} 
    \HeatMapProcedure{k}, \VoxelFlow{k} = \ModMVFB(\HeatMapBackHistoryBatchAll, \HeatMapProcedure{k-1}, \VoxelFlow{k-1}, \ScalingFactor). 
    \label{Eq.MVFB} 
\end{equation}
When $k=1$, $\HeatMapProcedure{0}$ and $\VoxelFlow{0}$ are set to zero.

On the other hand, the routing module is designed to dynamically balance the activation of each MVFB block, enabling adaptive selection according to the input variability. 
Contingent on a lightweight DNN, the routing module is 
optimized using Differentiable Bernoulli Sampling (DBS),  
to prevent the routing module from converging to trivial solutions (e.g., consistently activating or bypassing specific blocks). Specifically, DBS incorporates Gumbel-Softmax \cite{jang2022categorical} to determine the selection $\RoutingVector \in \{0,1\}$ of $k$-th MVFB through a stochastic classification task governed by $\RoutingProb$ as:  

\begin{equation}
    \RoutingVector = \frac{\expfunc{\RoutingProb + \GumbelNoise}}{\expfunc{\RoutingProb + \GumbelNoise} + \expfunc{2 - \RoutingProb - \GumbelNoise}},  
    \label{Gumbel}
\end{equation}  
where $\GumbelNoise \in \mathbb{R}$ follows the Gumbel(0,1) distribution. The temperature parameter $\Temperature$ starts with a high value to allow exploration of all possible paths and gradually decreases to approximate a one-hot distribution, ensuring effective and controllable routing.
To ensure the participation of DBS in gradient computation during end-to-end training, the Straight-Through Estimator (STE) \cite{bengio2013estimating}, which approximates the discrete sampling process in the backward pass, can be employed to further maintain compatibility with standard gradient descent optimization. 
  
In summary, the prediction process for the $k$-th MVFB is formulated as:
\begin{equation}
    \HeatMapProcedure{k}, \VoxelFlow{k} = 
        \begin{cases} 
            \ModMVFB(\HeatMapBackHistoryBatchAll, \HeatMapProcedure{k-1}, \VoxelFlow{k-1},\ScalingFactor), &\RoutingVector=1; \\
            \HeatMapProcedure{k-1}, \VoxelFlow{k-1}, &\RoutingVector=0.
        \end{cases}
    \label{Eq.DMVFN}
\end{equation}

To enhance the video prediction model's capacity for capturing dynamic information in traffic flow scenarios within the original training framework, we combine the standard $\ell_1$ loss, which controls the contribution of each stage and is regulated by a discount factor $\DiscountedRate$, with the VGG loss $\mathcal{L}_\text{Vgg}$ \cite{simonyan2015very} with a weight $\WeightLossVgg$. 
The VGG loss revolves around leveraging the feature extraction capabilities of pre-trained VGG networks to quantify perceptual differences between images. 
Mathematically, 
\begin{equation}
    \LossDMVFN = \sum\nolimits_{k=1}^{K} \DiscountedRate^{K-k} \ell_1(\HeatMapProcedureGT{t+\IntervalOverall}, \HeatMapProcedure{k}) + \WeightLossVgg\mathcal{L}_\text{Vgg}, 
    \label{Eq.LossDMVFN}
\end{equation}
where  
$
    \mathcal{L}_\text{Vgg} =\sum\limits_{m=1}^{M}\gamma_{m}\sum\limits_{h,w,c=1}^{\Height_m,\Width_m,\NumClass_m} \frac{1}{ \Height_m\Width_m\NumClass_m }\big( \phi_m(\HeatMapBackNext)_{h, w, c} - \phi_m(\HeatMapProcedure{t+\IntervalOverall})_{h, w, c} \big)^2 $. 
Here, $M=5$ indicates the number of VGG layers we chose in the off-the-shelf VGG-19 network \cite{simonyan2015very}. At the $m$-th layer, $\phi_m(\zeta)$ refers to the feature representation of input $\zeta$ and contributes to total loss with corresponding weight $\gamma_{m}$, and $\phi_m(\zeta)_{h, w, c}$ specifies the value of the feature map at the $h$-th row, $w$-th column, and $c$-th channel for the input $\zeta$ \cite{simonyan2015very}. $\Height_m$, $\Width_m$, and $\NumClass_m$ represent the height, width and channel count of the feature map at the $m$-th layer, respectively. 

\subsubsection{Motion-aware Affine Transformation (MAT)} 
Building upon the foundational work of \cite{wu2020motionnet}, $\ModMAT$ computes the motion prediction flow $\HeatMapFlow$, which explicitly encodes relative positional shifts between $\HeatMapBack$ and $\HeatMapProcedure{\ReceiveTime}$. 
\begin{equation}  
    \HeatMapFlow = \Phi_\text{MAT} (\HeatMapBack, \HeatMapProcedure{\ReceiveTime}),  
    \label{Eq.MAT}  
\end{equation} 

As depicted in Fig. \ref{Fig.DPP}, the MotionNet for $\ModMAT$ consists of two primary components: a Spatial-Temporal Pyramid Network (STPN) and a motion head, implemented by a two-layer 2D convolution module. The STPN is designed to extract multi-scale spatio-temporal features through its Spatial-Temporal Convolution (STC) block. 
The STC integrates standard 2D convolutions with a pseudo-1D convolution, which serves as a degenerate 3D convolution with kernel size $T_m \times 1 \times 1$, where $\{T_m\}_{m=1,2,3}$ corresponds to the temporal dimension, enabling efficient feature extraction across both spatial and temporal dimensions. 
Spatially, the STPN computes feature maps at multiple scales with a scaling factor of $2$, while temporally, it progressively reduces the temporal resolution to capture hierarchical temporal semantics. 
Following this, global temporal pooling, and a feature decoder with lateral connections and upsample layers are employed to aggregate and refine the extracted temporal features, ensuring robust motion representation. 

To precisely estimate the motion flow $\HeatMapFlow$, the loss function for $\ModMAT$ is defined using the smooth $\ell_1$ loss as:
\begin{align}
    &\mathcal{L}_\text{MAT} = \label{Eq.lossMAT} \\ &\norm{\sum_{k}\sum_{(x,y), (x',y')\in o_k} f_{A}\left( f_\Delta(\HeatMapAPCGT{(x,y)}{t}, \HeatMapAPCGT{(x',y')}{\ReceiveTime}) \right) - \HeatMapFlow}, \nonumber
\end{align}   
where 
$
    f_\Delta(\HeatMapAPCGT{(x,y)}{t}, \HeatMapAPCGT{(x',y')}{\ReceiveTime}) \in \mathbb{R}^{2} 
$ 
represents the aggregated motion (i.e., $(\Delta x, \Delta y) = (x',y') - (x,y)$) of object $k$ within instance $o_k$ over the interval $[t, \ReceiveTime]$, which is derived through grid-level comparisons between the Ground-Truth (GT) heatmaps $\HeatMapGT$ and $\HeatMapGTReceive$. 
The operator $f_{A}(\cdot)$ indicates a simple affine operation to map the 
increment $(\Delta x, \Delta y)$ to the $x$-th column, $y$-th row into a $H\times W$ matrix. 
Subsequently, the transformation of the semantic feature $\PseuBack$ can be directly performed with the help of motion flow $\HeatMapFlow$ as
\begin{equation}
    \PseuBackProcessed(x, y) = \PseuBack\left[x + \HeatMapFlow(x,y,0), y + \HeatMapFlow(x,y,1)\right]. 
    \label{Eq.affine}
\end{equation}

\subsection{AoIm-based Pragmatic Communications}
To incorporate driving-related information within the PragComm procedure, we initiate by generating the request map $\RequestMapEgo$. 
Given the inherent ambiguity of relying solely on navigation information $\Navigation$ \cite{Jaeger2023ICCV}, the request map is constructed as a Gaussian distribution centered on the nearest waypoint $(W_x, W_y)$ within prior waypoint plan $\WaypointsLast$, inspired by \cite{SCOUT}. The formulation is given by:
\begin{equation} 
    \RequestMapEgo(x,y) = \dfrac{1}{\FocusRadius\sqrt{2\pi}} \exp\left(-\dfrac{(x-W_x)^2+(y-W_y)^2}{2\FocusRadius^2}\right), \label{Eq.RequestMap} 
\end{equation}
where $\FocusRadius$, termed the \emph{Focus Radius}, is a hyperparameter controlling the width of the Gaussian distribution.

With the assistance of DPP, we further emphasize the dynamic information during message packing by computing $\ConfidenceDifference(x,y) = \abs{\ConfidenceMapGenerator(\HeatMapProcedure{\ReceiveTime})(x,y) - \ConfidenceMapBack(x,y)}$ as an alert signal, which has been proven to be practical in prior works \cite{lei2022latency}. The message $\Msg$ is then packed as:
\begin{equation} 
    \Msg = \PseuBackProcessed \times \PassingMask,
    \label{Eq.MessagePacking_APC} 
\end{equation}
where $\PassingMask = \OneMask{\Maximum{\RequestMapEgo \odot \ConfidenceMapGenerator(\HeatMapProcedure{\ReceiveTime}), \ConfidenceDifference/\LatencySteps} \geq \ThresholdCollab}$. Subsequently, the information delivery, fusion and decision-making procedures can be conducted as in Section \ref{sec:fusion}. In summary, Select2Drive can be executed as in Algorithm \ref{algo:Overall}.

\subsection{Training Methods}
In order to train the DNNs in Select2Drive, we assume the existence of a dataset $\DistributionDataset = \{\xi_{k}\}_{k=0 \dots N}$, which comprises trajectories $\xi_{k} = \{(\SensorEgo, \state, \WaypointsGT)\}_{t=0 \dots T}$ representing sequences of state-action pairs, with actions $\WaypointsGT = \PolicyExpert(\state)$ derived from an expert policy $\PolicyExpert$, where the real state $\state = (\HeatMapGT, \BoundingBoxGT, \Navigation)$. The training process is structured around two interconnected parts (i.e., the perception-related DNN and the planning policy). 
For the former part, a $\ParamPerception$-parameterized DNN $\ModPerception$, which encompasses the encoder $\ModEncoder$, decoder $\ModDecoder$, fuser $\ModFuse$ as well as the incorporated intermediate DNNs, especially DMVFN and MAT in DPP, is learned through minimizing the PointPillar perception loss through supervised learning,
\begin{equation}
    \underset{\ParamPerception}{\text{min}}\ \mathcal{L}(\ParamPerception) = 
    \mathbb{E}_{(\SensorEgo,\state)\in \DistributionDataset}[(\stateEstimate - \state)^2] 
    + \LossDMVFN + \LossMAT, 
    \label{Loss.percep}
\end{equation}
where $\stateEstimate = (\HeatMapFused, \BoundingBoxFused, \Navigation)$ represents the estimated state. 

On the other hand, the latter planning policy DNN $\ModPlan$ parameterized by $\ParamPlan$ is trained using IL to minimize the $l_2$-norm deviation \cite{hu2020collaborative} between the low-level planning strategies and the expert policy $\PolicyExpert$ as:   
\begin{equation}
    \underset{\ParamPlan}{\text{min}}\ \mathcal{L}(\ParamPlan) = 
    \mathbb{E}_{(\state,\WaypointsGT)\in \DistributionDataset}[(\Waypoints - \WaypointsGT)^2] 
\end{equation}  
where $\Waypoints = \ModPlan(\stateEstimate, \PseuFused)$ represents the waypoint plan using the estimated state \footnote{The occupancy map $\OccupancyMapEgo$ is derived through a non-trainable suppression mechanism and rasterization process applied to $\stateEstimate$.} $\stateEstimate$ along with fused semantic features $\PseuFused$ given by the perception-related DNN with converged parameters $\ParamPerception$. 
Since the optimization objective of $\ModPlan$ differs from that of $\ModPerception$, the planner is trained for the closed-loop task using features from the converged perception model.

\begin{algorithm}[t]
    \caption{Select2Drive} \label{algo:Overall}
    \SetAlgoRefName{1}
    \KwIn{Raw sensor data and last planned waypoints $\{\SensorEgo, \WaypointsLast \}_{\AllAgents}$ of ego $i$ and its neighboring agents $j\in\NAgents$,}
    \KwOut{Next driving action for each agent $\{\DriveAction\}_{\AllAgents}$}
    \BlankLine
    \For{each agent $i$}{ 
        Generate intermediate semantic features $\PseuEgo$ along with solo-perception results $\HeatMapEgo, \BoundingBoxEgo$ based on $\SensorEgo$ using Eqs. \eqref{Eq.extraction}\eqref{Eq.decoder}\;
        Exchange request map $\RequestMapEgo$ based on prior driving plan $\WaypointsLast$ using Eq. \eqref{Eq.RequestMap} and estimate latency $\LatencyOverallEst$ of sending message to neighbor $j \in \NAgents$\;
        \For{neighboring agent $\NAgents$}{
            $\LatencySteps \leftarrow \lfloor \LatencyOverallEst / \tau \rfloor$\; 
            Predict future heatmap $\HeatMapProcedure{\ReceiveTime}$ based on historical information $\HeatMapBackHistoryBatchAll$ through $\LatencySteps$ iterations of DMVFN in Eq. \eqref{Eq.DMVFN}\;
            Extract motion flow $\HeatMapFlow$ between $\HeatMapBack$ and $\HeatMapProcedure{\ReceiveTime}$ with MAT in Eq. \eqref{Eq.MAT} \;
            Apply affine approximation $\HeatMapFlow$ on semantic features $\PseuBack$ to estimate high-level semantic information $\PseuBackProcessed$ with Eq. \eqref{Eq.affine} ; \algorithmiccomment{DPP} \\
            Send packed Message $\Msg$ based on $\PseuBackProcessed$ and confidence map $\ConfidenceMapEgo$ generated with Eq. \eqref{Eq.confidence} using Eq.~\eqref{Eq.MessagePacking_APC} ;\algorithmiccomment{APC}\\
        }
        Fuse received message $\MsgAll$ and ego information $\PseuEgo$ to obtain collaborated semantics $\PseuFused$ using Eq. \eqref{Eq.fusion_MHA}\;
        Generate next driving plan $\Waypoints$ and make driving decision $\DriveAction$ based on $\PseuFused$ using Eqs. \eqref{Eq.decode_fused}\eqref{Eq.control}\;
    }
    \Return Next action $\{\DriveAction\}_{\AllAgents}$
\end{algorithm}\DecMargin{1em}

\begin{table}
    \centering
    \caption{Mainly used parameters in this paper.}
    \label{tab.params}
    \begin{tabular}{|c|m{2.5cm}|}
    \toprule
    \multicolumn{1}{|c|}{\textbf{Parameter}} & \textbf{Value} \\ \midrule
    \multicolumn{2}{|c|}{\textbf{DSRC-based transmission}} \\ \hline
    \multicolumn{1}{|c|}{Interval $\IntervalSync$ for SCHI and CCHI} & $50$ ms \\ \hline
    \multicolumn{1}{|c|}{Fixed Decision Interval $\IntervalOverall$} & $100$ ms \\ \hline
    \multicolumn{1}{|c|}{Allocated Bandwidth $\BandWidthSingle$} & $1\sim20$ MHz \\ \hline
    \multicolumn{1}{|c|}{Transmit Power $\TransPower$} & $23$ dBm \\ \hline
    \multicolumn{1}{|c|}{Power of Noise $\TransNoise$} & {$U(-95, -110)$} dBm \\ \hline
    \multicolumn{1}{|c|}{Carrier Frequency $\CarrierFrequency$} & $5.9$ GHz \\ \hline
    \multicolumn{2}{|c|}{\textbf{C-V2X-based transmission(ms)}} \\ \hline
    \multicolumn{1}{|c|}{Fixed transmission Latency $\LatencyPropagation+\LatencyNetwork$} & $0\sim600$\\ \hline
    \multicolumn{2}{|c|}{\textbf{Shared Latency-related parameters}} \\ \hline
    \multicolumn{1}{|c|}{Packet Loss} & $5\%$\\ \hline
    \multicolumn{1}{|c|}{Asynchronous latency $\LatencyAsync$} & {$U(-100, 100)$} ms\\ \hline
    \multicolumn{1}{|c|}{Queueing latency $\LatencyQueue$} & {$U(0, 50)$} ms\\ \hline
    \multicolumn{1}{|c|}{Semantic Extraction Time $\LatencyExt$} & {$U(40, 50)$} ms\\ \hline
    \multicolumn{1}{|c|}{Decision-Making Time $\LatencyDecision$} & {$U(20, 30)$} ms\\ \hline
    \multicolumn{2}{|c|}{\textbf{Hyperparameters}} \\ \hline
    \multicolumn{1}{|c|}{Height, Width, Channel $\{\Height, \Width, \Channel\}$} & {$[192, 576, 64]$} \\ \hline
    \multicolumn{1}{|c|}{Request Map Threshold $\ThresholdCollab$} & $0.05$ \\ \hline
    \multicolumn{1}{|c|}{Focus Radius $\FocusRadius$} & $15$ m \\ \hline
    \multicolumn{1}{|c|}{Number of frames for planning $\LenDrivingMemory$} & $5$ \\ \hline
    \multicolumn{1}{|c|}{Number of waypoints to plan $\LenDrivingTrajectory$} & $10$ \\ \hline
    \multicolumn{1}{|c|}{Scaling factors $\{\ScalingFactor\}_{k=1}^9$} & \begin{tabular}[c]{@{}l@{}}$[4, 4, 4, 2, 2, 2,$ \\ $1, 1, 1]$\end{tabular} \\ \hline
    \multicolumn{1}{|c|}{Discount factor $\DiscountedRate$, VGG weight $\WeightLossVgg$} & $0.8, 0.5$ \\ \hline
    \multicolumn{1}{|c|}{Index $\{m\}_M$ of VGG Layers} & $[2,\,7,\,12,\,21,\,30]$ \\ \hline
    \multicolumn{1}{|c|}{Corresponding weights $\{\gamma_{m}\}_M$} & \begin{tabular}[c]{@{}l@{}}$[0.38,\,0.21,\,0.27,$\\$\,0.18,\,6.67]$\end{tabular}  \\ \hline
    \multicolumn{1}{|c|}{Temporal factors in STPN $T_1, T_2, T_3$} & $[2, 2, 1]$ \\ \bottomrule
    \end{tabular}
    \vspace{-4mm}
\end{table}

\section{Experimental Results and Discussions}\label{sec5}
\subsection{Experimental Settings}
In this section, we evaluate the performance of Select2Drive in a high-fidelity environment based on CARLA simulator, which facilitates sensor rendering and the computation of physics-based updates to the world state. 
It adheres to the ASAM OpenDRIVE standard \cite{OpenDrive} for defining road networks and urban environments. 
{Table \ref{tab.params} outlines the principal experimental parameters, with communications-related parameters primarily mentioned in \cite{8356112}. 
The values of $\LatencyExt$ and $\LatencyDecision$ are obtained from Table \ref{tab.modules}. Specifically, $\LatencyExt$ denotes the aggregated latency of $\ModEncoder$ and $\ModProcess$ with an average of $43.71$ ms, 
whereas $\LatencyDecision$ represents the cumulative delay of $\ModDecoder$, $\ModFuse$, and $\ModPlan$, averaging $24.34$ ms. 
Furthermore, the value of the queueing latency $\LatencyQueue$ corresponds to {$30.42$} ms under conditions with $5$ or more communicable agents, assuming a DSRC channel throughput of $20$ Mbps and an arrival rate of $10$ Hz modeled as an M/M/1 model \cite{hwang2020communication}.}

The overall latency $\LatencyOverall$ is simulated separately with assumed bandwidth constraints in DSRC \cite{IEEE802} and C-V2X \cite{7992934} as in Section \ref{sec.latency}. 
Since DSRC-based transmission encounters hidden node issues, which can lead to packet collisions \cite{Sedar2023V2X}, it is modeled by constraining $\LatencyPropagation$ with limited bandwidth, while $\LatencyNetwork = 0$ due to its direct communication nature. 
In contrast, in C-V2X, the impact of $\LatencyNetwork$ is more pronounced due to the possible handover procedures, and $\LatencyTrans$ {fluctuates within a bounded range.} 
{Motivated by the practice in \cite{wang2024motion}, 
we simulate packet loss by applying random dropout on the transmitted message $\Msg$, and model jitter $\LatencyAsync$ with varying variance levels.
Specifically, when packet loss occurs, the features received by the ego vehicle are replaced with Gaussian noise, whereas jitter shifts the receive timestamp of semantic features, causing them to arrive earlier or later than expected.} 

As the V2X-AD framework naturally divides into perception and subsequent driving tasks, we evaluate our proposed method across two distinct stages. 

\begin{figure*}[tbp]
    \centering
    \subfloat[][Pedestrian crossing]{\includegraphics[width=0.32\linewidth]{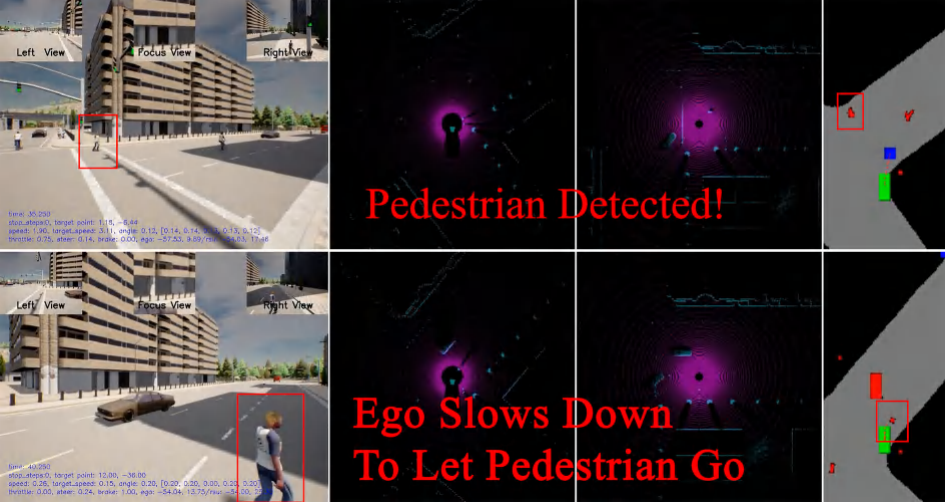}} \hspace{.01em}
    \subfloat[][Bicycle Overtaking]{\includegraphics[width=0.32\linewidth]{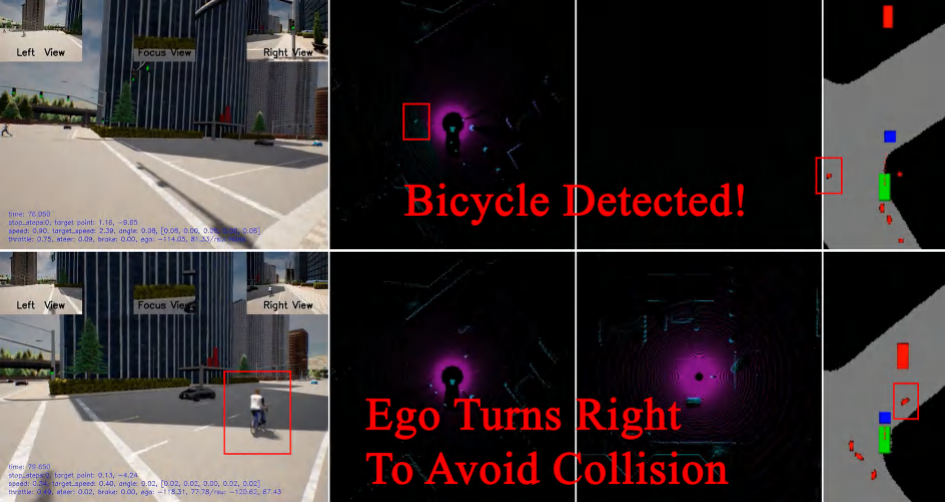}} \hspace{.01em}
    \subfloat[][Pedestrian crossing from blind spot]{\includegraphics[width=0.32\linewidth]{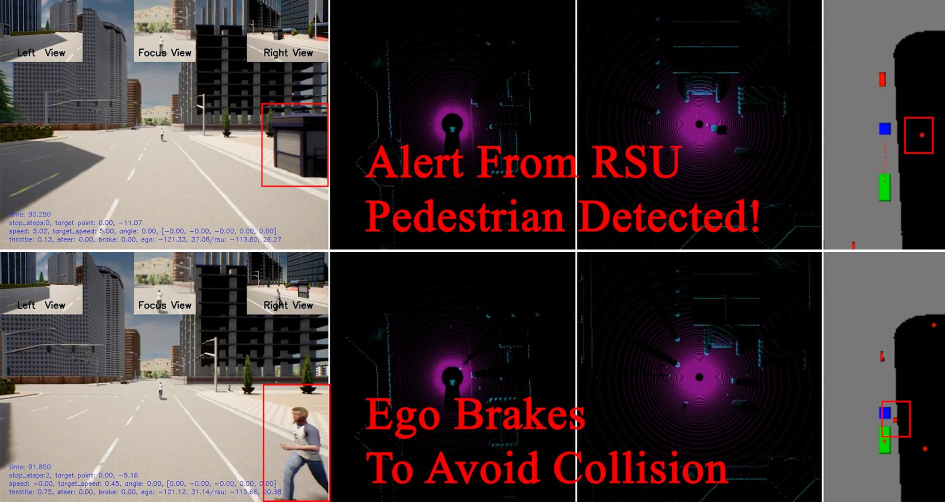}}
    \caption{Visualization for closed-loop driving upon several accident scenarios. 
    The visualization delineates the ego vehicle's position with a \textcolor{green}{green} box, the planned trajectory with a \textcolor{red}{red} dot, detected obstacles as red squares, and the next waypoint along the route as a \textcolor{blue}{blue} square.}
    \label{Fig.closeloop}
    \vspace{-2mm}
\end{figure*}

\begin{table*}[htbp]
    \centering
    \caption{Closed-loop driving performance. The approach with the best average driving score is highlighted in \textbf{bold}, while the second-best and the third-best are marked with \textit{italics} and \uline{underlines}, respectively.} 
    \begin{threeparttable}
    \begin{tabular}{|m{2.2cm}|cccccccc|}
    \toprule
    \textbf{Method} & {\begin{tabular}[c]{@{}c@{}}\textbf{Avg.} \\ \textbf{Driving} \\ \textbf{Scores}$\uparrow$\end{tabular}} & {\begin{tabular}[c]{@{}c@{}}\textbf{Avg. Route} \\ \textbf{Completion} \\ \textbf{Rate} (\%)$\uparrow$\end{tabular}} & {\begin{tabular}[c]{@{}c@{}}\textbf{Avg.} \\ \textbf{Infraction} \\ \textbf{Penalty}$\downarrow$\end{tabular}} & {\begin{tabular}[c]{@{}c@{}}\textbf{Collisions} \\ \textbf{With} \\ \textbf{Pedestrians}$\downarrow$\end{tabular}} & {\begin{tabular}[c]{@{}c@{}}\textbf{Collisions} \\ \textbf{With} \\ \textbf{Vehicles}$\downarrow$\end{tabular}} & {\begin{tabular}[c]{@{}c@{}}\textbf{Collisions} \\ \textbf{With} \\ \textbf{Layout}$\downarrow$\end{tabular}} & {\begin{tabular}[c]{@{}c@{}}\textbf{Off-road} \\ \textbf{Infractions}$\downarrow$\end{tabular}} & \begin{tabular}[c]{@{}c@{}}\textbf{Mean}\\ \textbf{Speed}$\uparrow$\end{tabular} \\ \midrule
    \multicolumn{9}{|c|}{\textbf{No Communications}} \\ \cline{1-9}
    {\underline{Interfuser} \cite{shao2022interfuser}} & $35.372$ & $79.254$ & $0.434$ & $0.052$ & $0.492$ & $0.568$ & $0.223$ & ${0.586}$ \\ \cline{1-9}
    {\textit{TCP \cite{wu2022tcp}}} & $38.214$ & $50.526$ & $0.817$ & $0.029$ & $0.079$ & $0.069$ & $0.004$ & $1.066$ \\ \cline{1-9}
    {\textbf{No Fusion}} & $38.481$ & $84.732$ & $0.432$ & $0.109$ & $0.379$ & $0.603$ & $0.105$ & $0.569$ \\ \midrule
    \multicolumn{9}{|c|}{\textbf{Bandwidth =} $20$ \textbf{MHz (uniform latency =} $0$ \textbf{ms)}} \\ \cline{1-9}
    {When2Com \cite{liu2020when2com}} & {$30.571$} & {$41.840$} & {$0.646$} & {$0.028$} & {$0.923$} & {$0.450$} & {$0.416$} & $0.218$ \\ \cline{1-9}
    {Where2Comm \cite{hu2022where2comm}} & {$35.811$} & {$82.266$} & {$0.394$} & {$0.156$} & {$0.390$} & {$0.393$} & {$0.115$} & $0.791$ \\ \cline{1-9}
    {Select2Col \cite{liu2024select2col}} & {$35.178$} & {$69.045$} & {$0.492$} & {$0.126$} & {$0.572$} & {$0.371$} & {$0.106$} & $0.442$ \\ \cline{1-9}
    {{\textit{SiCP} \cite{qu2024sicp}}} & $43.289$ & $80.159$ & $0.466$ & $0.111$ & $0.205$ & $0.852$ & $0.071$ & $1.082$ \\ \cline{1-9}
    {\uline{Select2Drive wo APC}} & $40.991$ & $82.535$ & $0.411$ & $0.148$ & $0.447$ & $0.404$ & $0.126$ & $0.978$ \\ \cline{1-9}
    {\textbf{Select2Drive}} & $46.904$ & $82.284$ & $0.446$ & $0.140$ & $0.270$ & $0.008$ & $0.083$ & $1.211$ \\ \cline{1-9}
    \multicolumn{9}{|c|}{\textbf{Bandwidth =} $10$ \textbf{MHz (uniform latency =} $100$ \textbf{ms)}} \\ \cline{1-9}
    {When2Com} & {$29.915$} & {$43.725$} & {$0.632$} & {$0.051$} & {$0.953$} & {$0.410$} & {$0.385$} & $0.257$ \\ \cline{1-9}
    {Where2Comm} & {$33.704$} & {$48.560$} & {$0.651$} & {$0.054$} & {$0.640$} & {$0.351$} & {$0.115$} & $0.668$ \\ \cline{1-9}
    {Select2Col} & {$29.794$} & {$70.232$} & {$0.414$} & {$0.189$} & {$0.516$} & {$0.348$} & {$0.118$} & $0.448$ \\ \cline{1-9}
    {{\textit{SiCP}}} & $41.849$ & $78.755$ & $0.418$ & $0.124$ & $0.215$ & $0.755$ & $0.079$ & $1.107$ \\ \cline{1-9}
    {\uline{Select2Drive wo APC}} & $40.725$ & $66.760$ & $0.496$ & $0.052$ & $0.469$ & $0.482$ & $0.142$ & $1.066$ \\ \cline{1-9}
    {\textbf{Select2Drive}} & $45.062$ & $81.157$ & $0.456$ & $0.117$ & $0.310$ & $0.376$ & $0.095$ & $0.976$ \\ \midrule
    \multicolumn{9}{|c|}{\textbf{Bandwidth =} $5$ \textbf{MHz (uniform latency =} $200$ \textbf{ms)}} \\ \cline{1-9}
    {When2Com} & {$27.204$} & {$38.392$} & {$0.652$} & {$0.043$} & {$1.114$} & {$0.377$} & {$0.514$} & $0.507$ \\ \cline{1-9}
    {Where2Comm} & {$31.976$} & {$51.161$} & {$0.527$} & {$0.119$} & {$0.514$} & {$0.399$} & {$0.188$} & $0.306$ \\ \cline{1-9}
    {Select2Col} & {$28.391$} & {$64.284$} & {$0.447$} & {$0.096$} & {$0.586$} & {$0.345$} & {$0.137$} & $0.486$ \\ \cline{1-9}
    {{\textit{SiCP}}} & $40.511$ & $66.415$ & $0.405$ & $0.132$ & $0.229$ & $0.795$ & $0.075$ & {$0.912$} \\ \cline{1-9}
    {\uline{Select2Drive wo APC}} & $38.853$ & $54.574$ & $0.627$ & $0.044$ & $0.626$ & $0.340$ & $0.331$ & $0.860$ \\ \cline{1-9}
    {\textbf{Select2Drive}} & $43.823$ & $70.588$ & $0.520$ & $0.088$ & $0.373$ & $0.497$ & $0.126$ & $1.211$ \\ \bottomrule
    \end{tabular}
        \begin{tablenotes}
            \footnotesize
            \item[1] {The experimental results present averaged measurements across $31$ independent routes, evaluated under varying seed parameters while maintaining fixed shared parameters as specified in Table \ref{tab.params}.}
            \normalsize
        \end{tablenotes}
    \end{threeparttable}
    \label{tab.driving}
    \vspace{-4mm}
\end{table*}

\begin{itemize} 
\item \emph{For planning policy}, we mainly simulate the closed-loop driving task through
online route completion tasks. 
%
%
All decision-making policies are pre-trained on V2Xverse \cite{liu2024towards}, while online tasks are tested on the 31 Town05 Short Routes in the CARLA Leaderboard \cite{carlaleaderboard} version 0.9.10, where the ego vehicle collaborates with the nearest agents (including vehicles and RSUs). Following \cite{jia2024bench2drive}, we employ three key evaluation metrics, including \emph{route completion rate}, \emph{infraction penalty}, and \emph{driving score}, as mentioned in Section \ref{sec3}.
Route completion rates quantify the agent's ability to successfully complete navigation tasks, calculated as the percentage of the planned route traversed. 
Infraction penalty evaluates traffic rule compliance through a geometric penalty function that accounts for both violation severity and frequency. 
Driving scores integrate the aforementioned factors along with collision rates, serving as a more comprehensive performance metric. Consequently, the ranking in the table primarily follows the driving score criterion.
\item \emph{For perception capability}, 
we leverage the V2Xverse \cite{liu2024towards} {and DAIR-V2X \cite{yu2022dair} datasets} for offline perception performance evaluation. 
The {former} dataset comprehensively incorporates RSUs compared to the widely used OPV2V dataset \cite{xu2022opv2v}, extending beyond vehicle-to-vehicle (V2V) communications, {while the latter is the latest real-world vehicle-infrastructure cooperative dataset. 
Notably, though some vehicles might not participate in the communications of perceived data to the ego vehicle, to account for their potential communications to others, we equally allocate the available bandwidth among all the ego's neighboring vehicles capable of communications.} 
For evaluation, consistent with \cite{liu2024select2col}, we adopt \emph{Average Precision (AP) at Intersection over Union (IoU)} thresholds of $0.3$, $0.5$, and $0.7$ for vehicles, two-wheeled vehicles, and pedestrians, denoted as AP30, AP50, and AP70. As for communication volume, we calculate it as $\log_2(\Height\times\Width\times\Channel\times\Vert\PassingMask\Vert_{1}\times $32/8$)$ \cite{hu2022where2comm}. 
To enhance clarity, the Composited AP is a weighted sum of AP30, AP50, and AP70, with respective weights of $0.3$, $0.3$, and $0.4$.
Also, to streamline representation, perception results for vehicles, bicycles, and pedestrians are merged with weights of $0.4$, $0.4$, and $0.2$ in Latency-induced scenarios shown in Fig. \ref{Fig.bandwidth} and Fig. \ref{Fig.packet}. 
To mitigate excessive oscillations in the curves caused by positioning noise, the weights become $0.8$, $0.1$, and $0.1$ in Positioning-induced scenarios shown in Fig. \ref{Fig.noise}.
{To quantitatively assess the statistical impact of latency fluctuations, we present performance curves accompanied by their corresponding confidence intervals. The solid line represents the mean values, while the shaded regions delineate the upper and lower bounds of the confidence interval.}
\end{itemize}
\emph{For baseline methodologies}, 
we reproduce When2Com \cite{liu2020when2com}, Where2Comm \cite{hu2022where2comm} and Select2Col \cite{liu2024select2col}, {as well as the State-Of-The-Art (SOTA) SiCP \cite{qu2024sicp}, as the baseline of the perception task.} 
Additionally, a no-fusion baseline is included to evaluate performance in the absence of collaborative mechanisms.
Meanwhile, the baselines in the driving task include an IL-based planner trained atop the collaborative perception methods above. 
Additionally, prominent single-agent end-to-end methodologies are also considered,  such as TCP \cite{wu2022tcp} and the SOTA Interfuser \cite{shao2022interfuser}. 

\subsection{{Quantitative} Results}
\subsubsection{Driving Task}
{Quantitatively}, as illustrated in Fig. \ref{Fig.closeloop}, the proposed approach effectively leverages PragComm-based driving-critical information for emergent obstacle perception and timely collision avoidance. 

Table \ref{tab.driving} demonstrates a marked performance enhancement of 
$8.35\%$ (resp., $3.62$)
in driving scores using the proposed methodology compared with the SOTA approach {SiCP}. 
Under bandwidth constraints, the collaborative driving paradigm maintains a 
{$8.18\%$ (resp., $3.31$)}
Driving scores advantage compared to the latest available single-agent SOTA TCP method \cite{wu2022tcp}. 
Notable gains can be observed for the road completion rate  ({$2.65\%$}
improvement) and infraction penalty ({$4.29\%$}
reduction). 
These results confirm the universal superiority and robustness of our method in real-world scenarios.
Ablation studies confirm the contribution of the APC methodology, yielding a $14.43\%$ 
improvement in the driving score and empirically validating our ``less is more" hypothesis. 
%
Conversely, latency-agnostic methods \cite{liu2020when2com, hu2022where2comm} exhibit significant performance degradation under high-latency conditions.

\begin{figure}[tbp]
    \centering
    \includegraphics[width=0.85\linewidth]{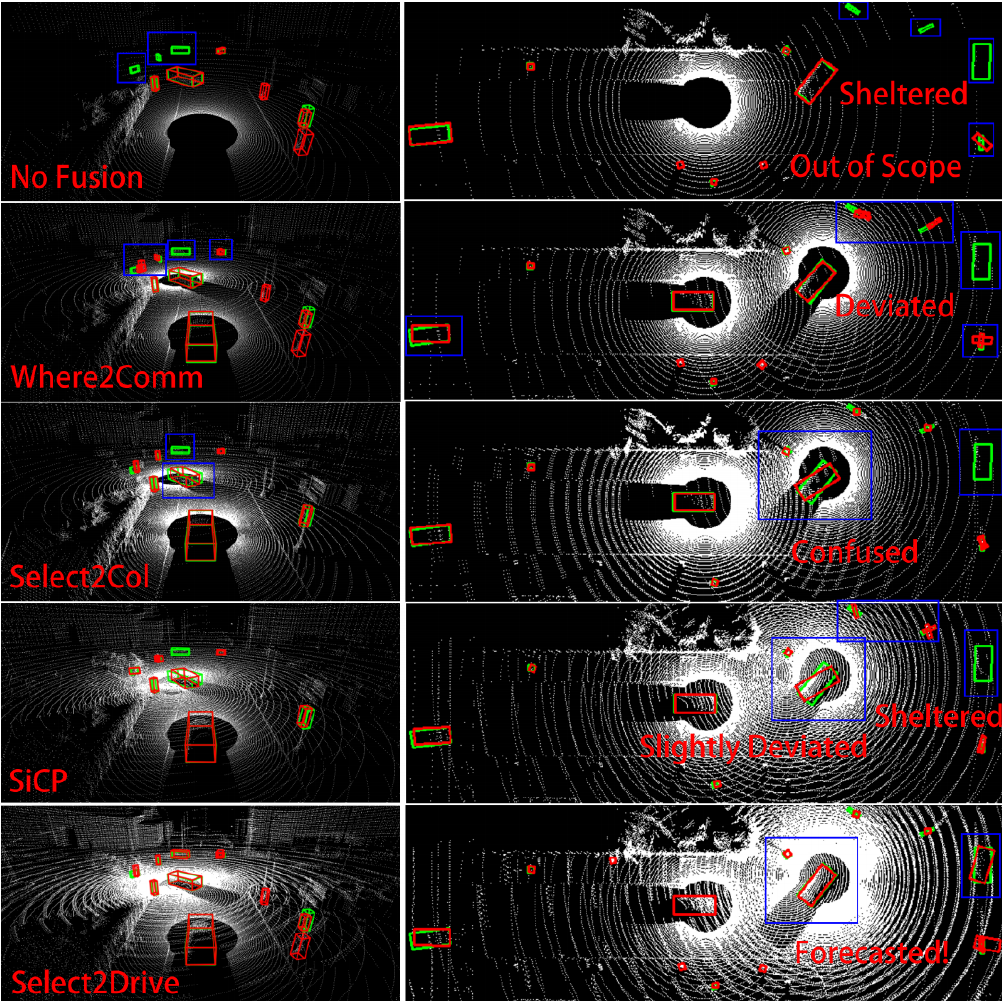}
    \caption{Visualization of collaborative perception in bandwidth-constrained ($5$ MHz) scenarios. The \textcolor{red}{red} box illustrates the ego vehicle's predicted positions for surrounding objects, whereas the \textcolor{green}{green} box indicates the GT positions of those objects.}
    \label{Fig.visualize}
    \vspace{-4mm}
\end{figure}

\begin{figure}[tbp]
	\centering
        \subfloat[][DSRC-based Transmission in V2Xverse.]{\includegraphics[width=0.5\linewidth]{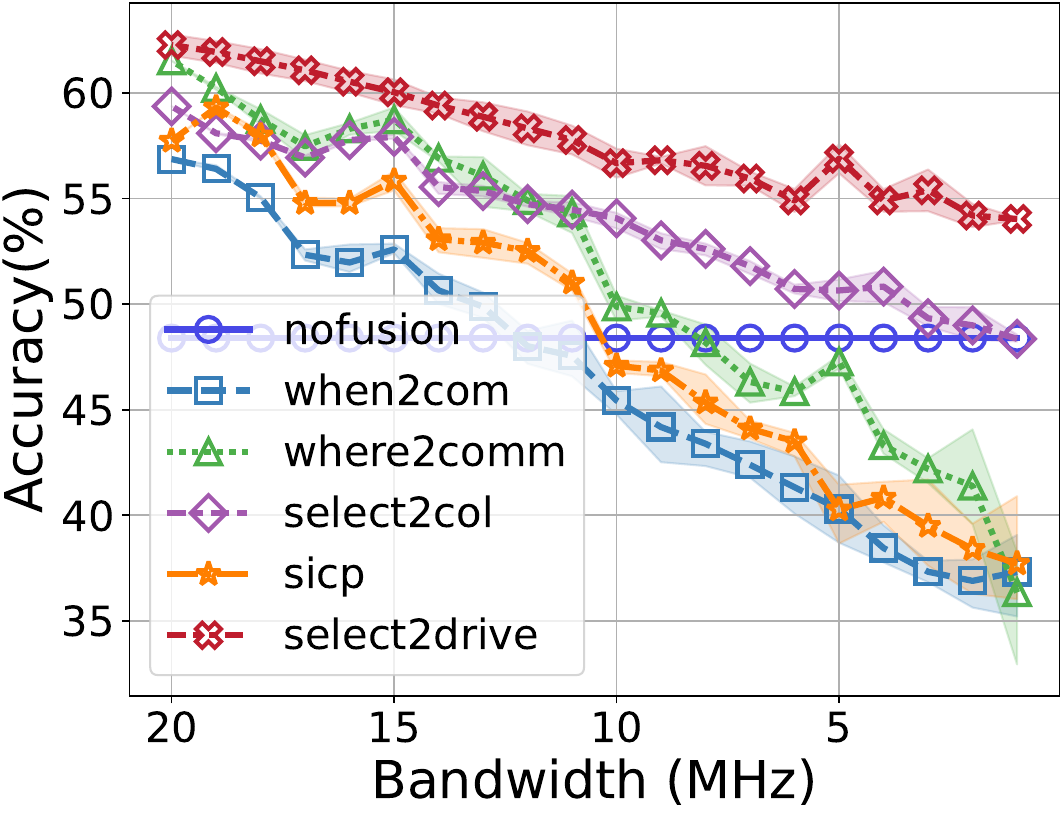}} 
        \subfloat[][C-V2X-based Transmission in V2Xverse.]{\includegraphics[width=0.5\linewidth]{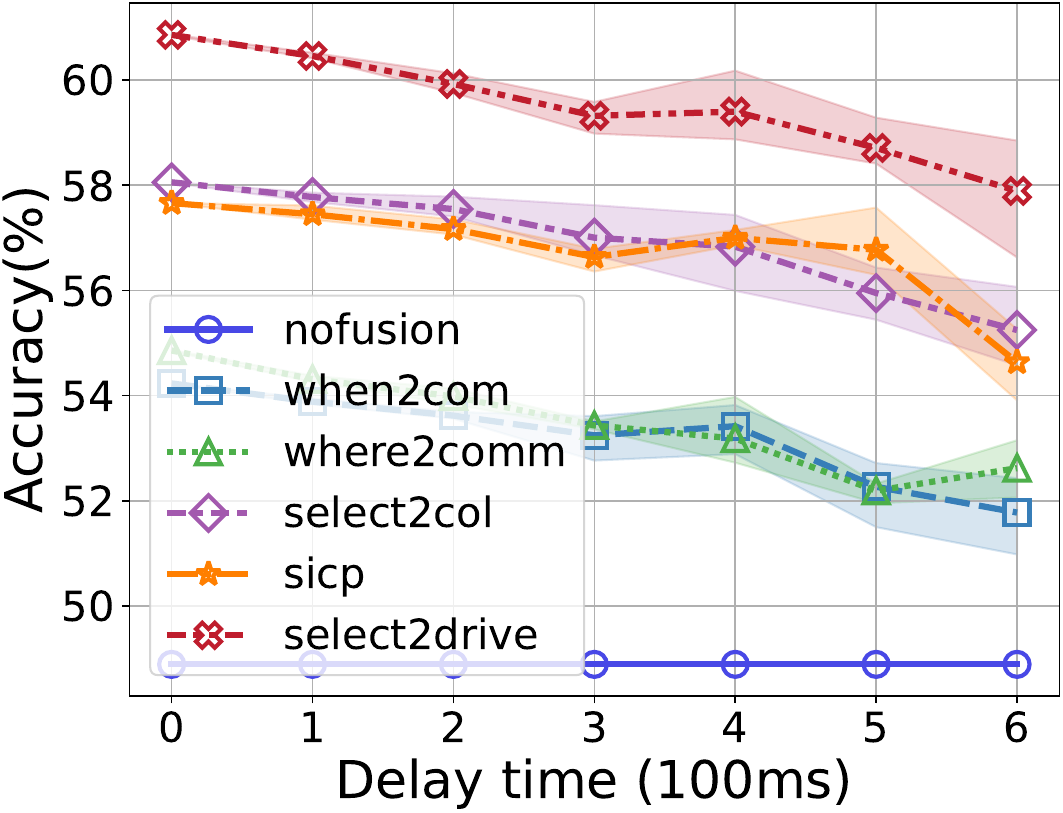}}
        \\
        \subfloat[][DSRC-based Transmission in DAIR-V2X.]{\includegraphics[width=0.5\linewidth]{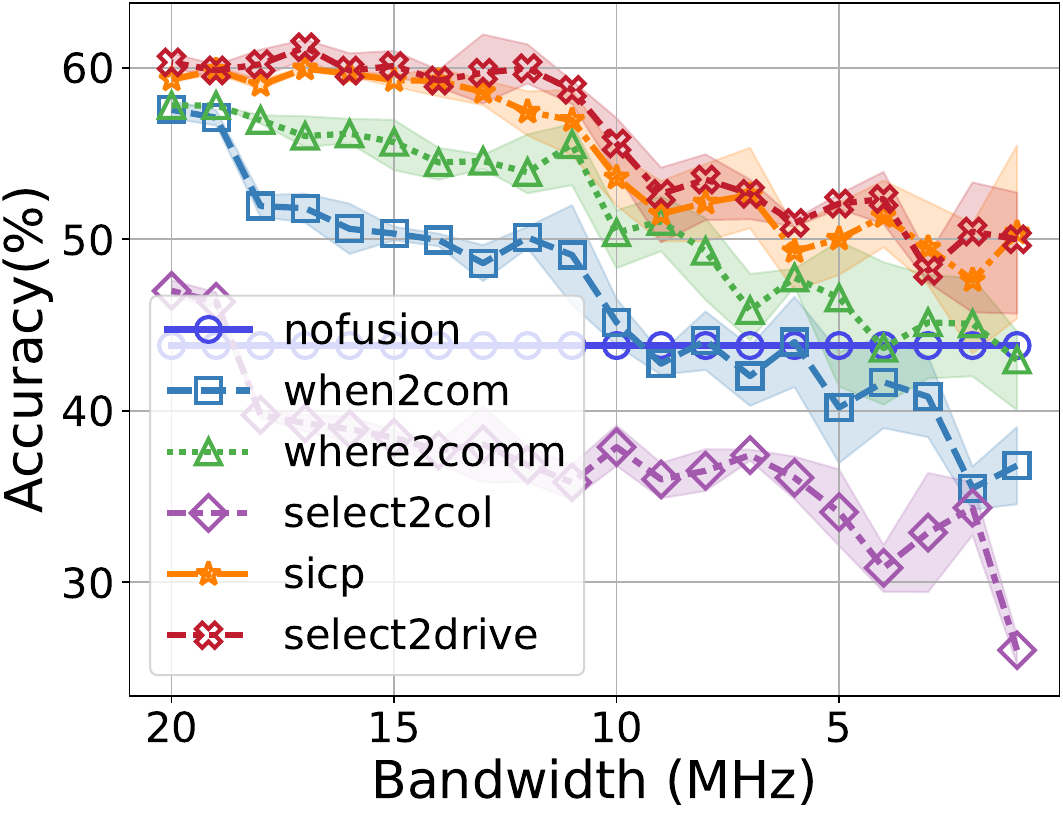}} 
        \subfloat[][C-V2X-based Transmission in  DAIR-V2X.]{\includegraphics[width=0.5\linewidth]{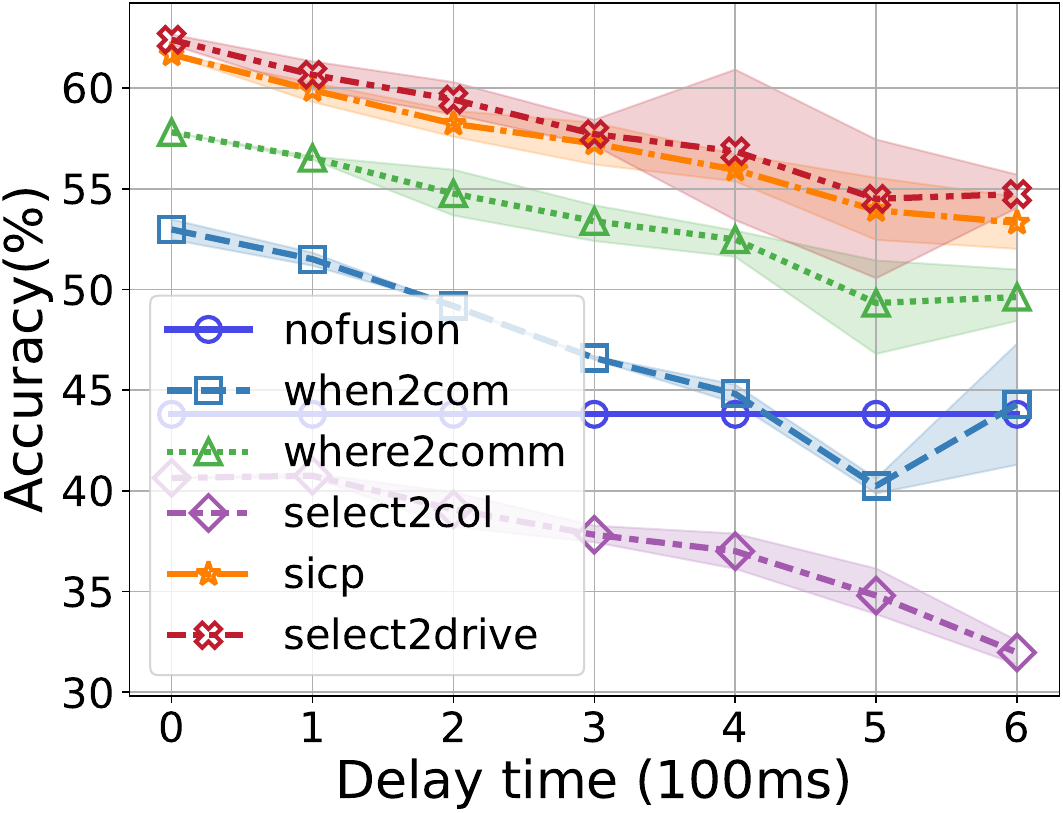}}
    \caption{Robustness of DPP to the communication constraints in the perception task.}
    \label{Fig.bandwidth}
    \vspace{-6mm}
\end{figure} 
\begin{figure*}[tbp]
	\centering
    \subfloat[Composited AP for $\NoiseRotation = 0$ in V2XVerse.]{\includegraphics[width=0.3 \linewidth]{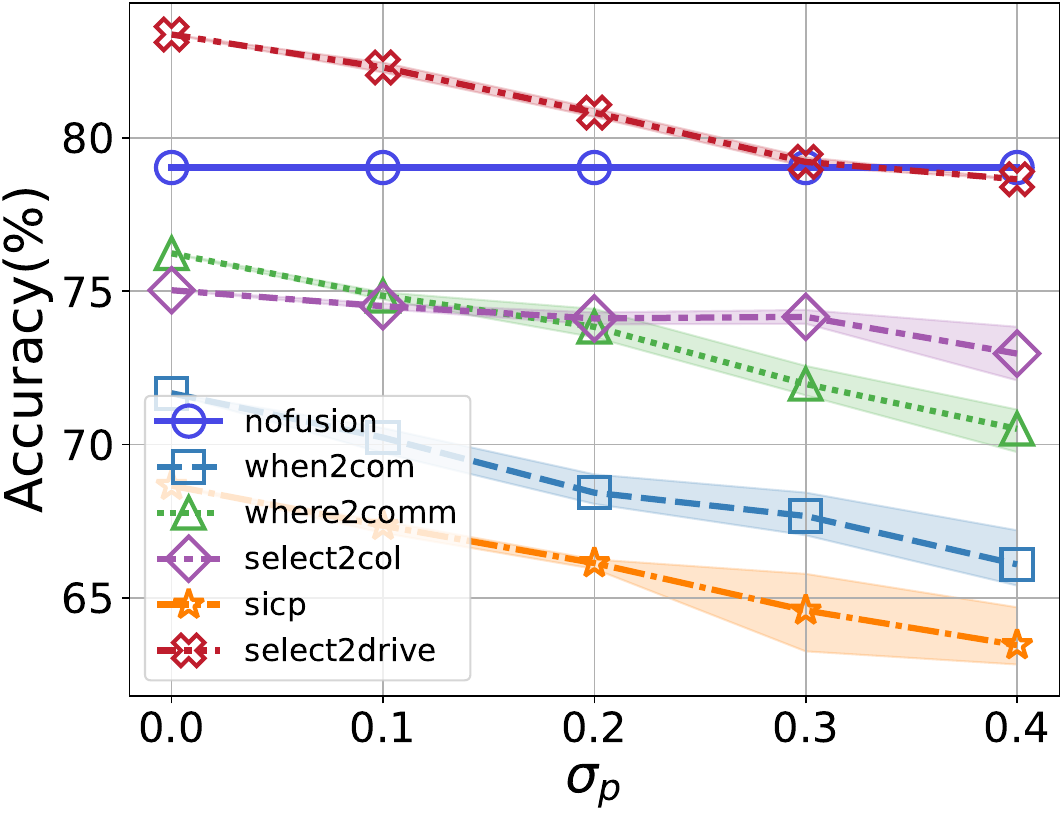}} 
    \subfloat[Composited AP for $\NoisePosition = 0$  in V2XVerse.]{\includegraphics[width=0.3 \linewidth]{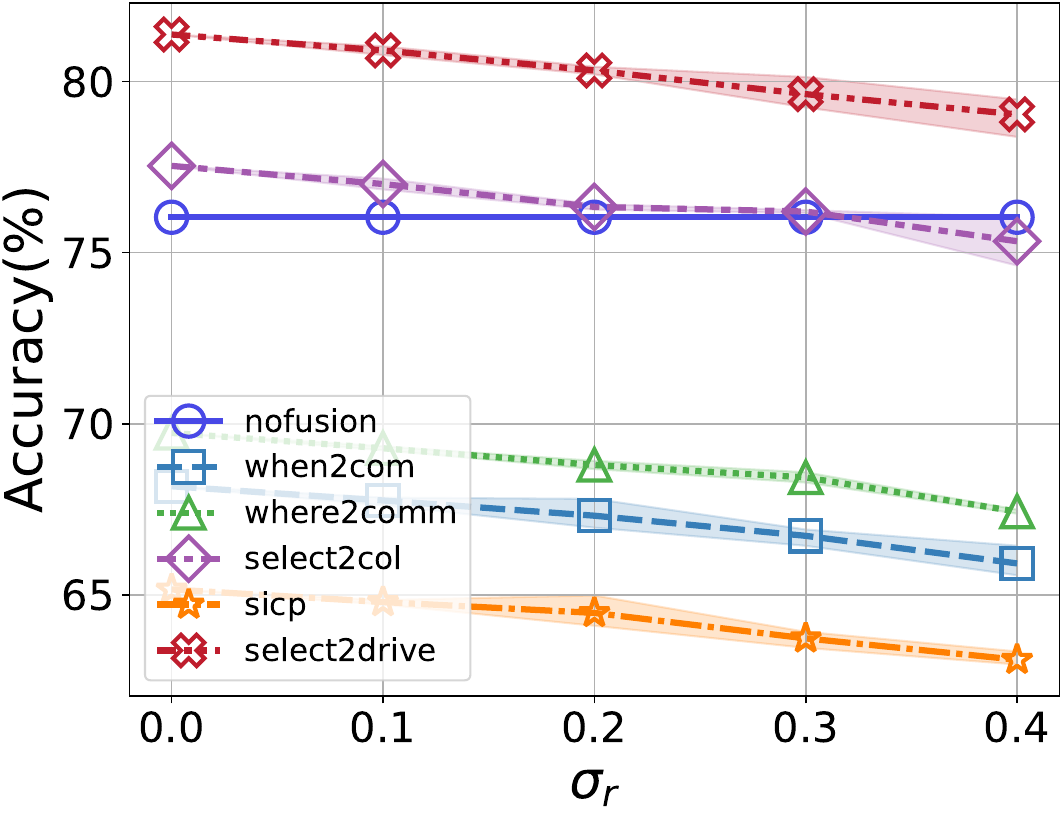}} 
    \subfloat[Composited AP in V2XVerse.]{\includegraphics[width=0.3 \linewidth]{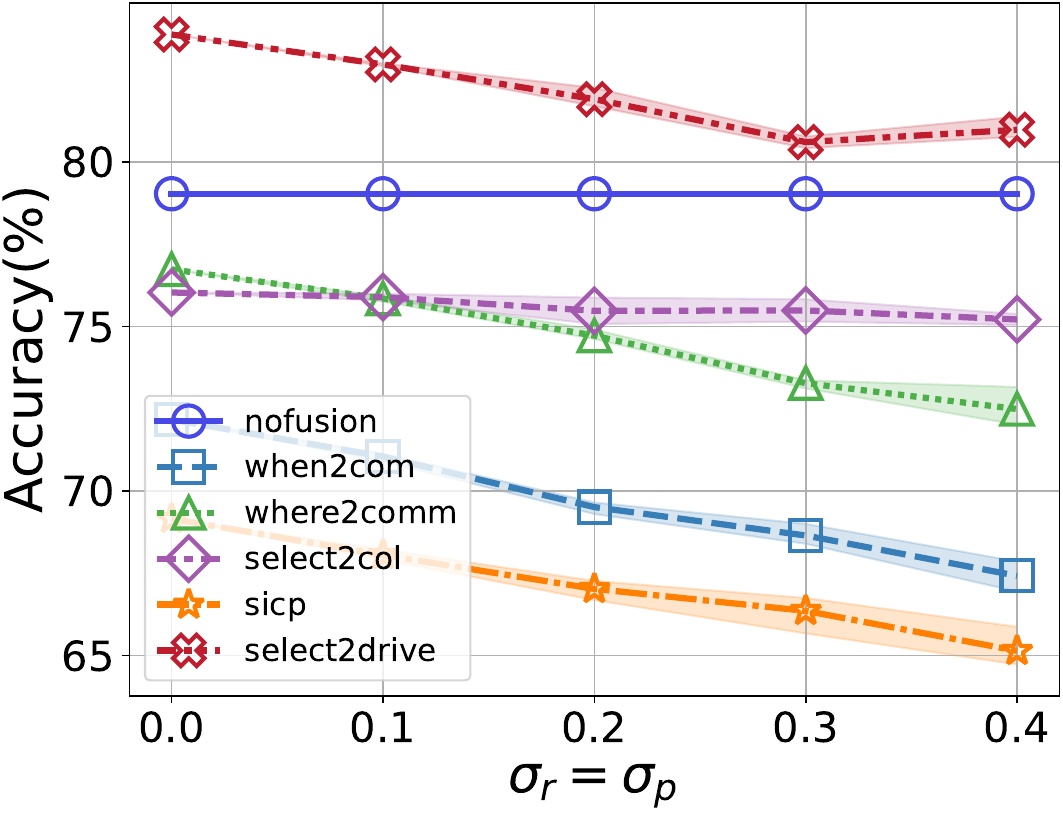}} \\
    \subfloat[Composited AP for $\NoiseRotation = 0$ in DAIR-V2X.]{\includegraphics[width=0.3 \linewidth]{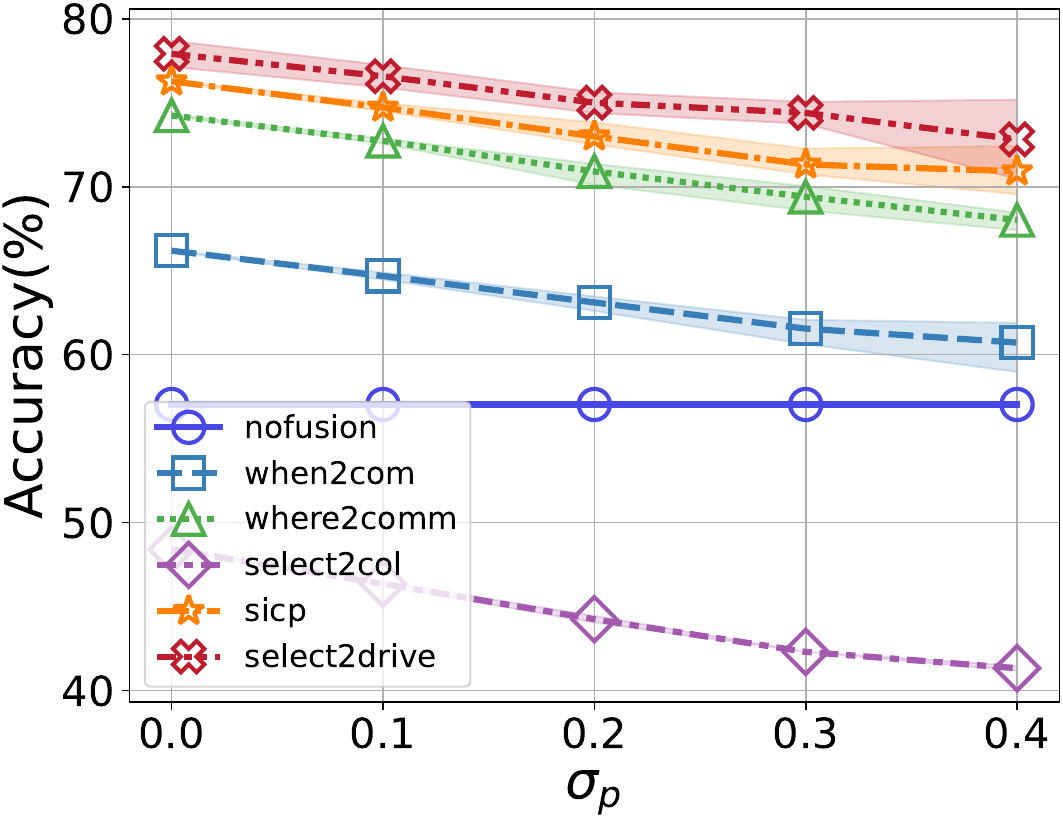}} 
    \subfloat[Composited AP for $\NoisePosition = 0$  in DAIR-V2X.]{\includegraphics[width=0.3 \linewidth]{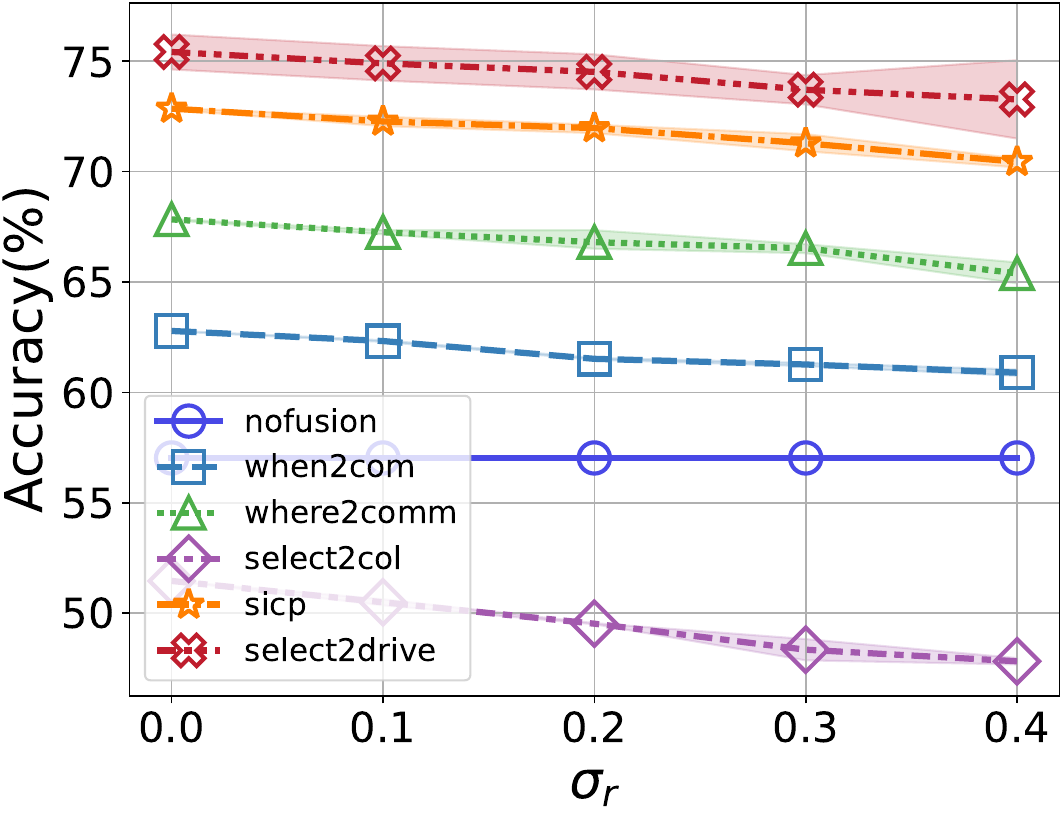}} 
    \subfloat[Composited AP in DAIR-V2X.]{\includegraphics[width=0.3\linewidth]{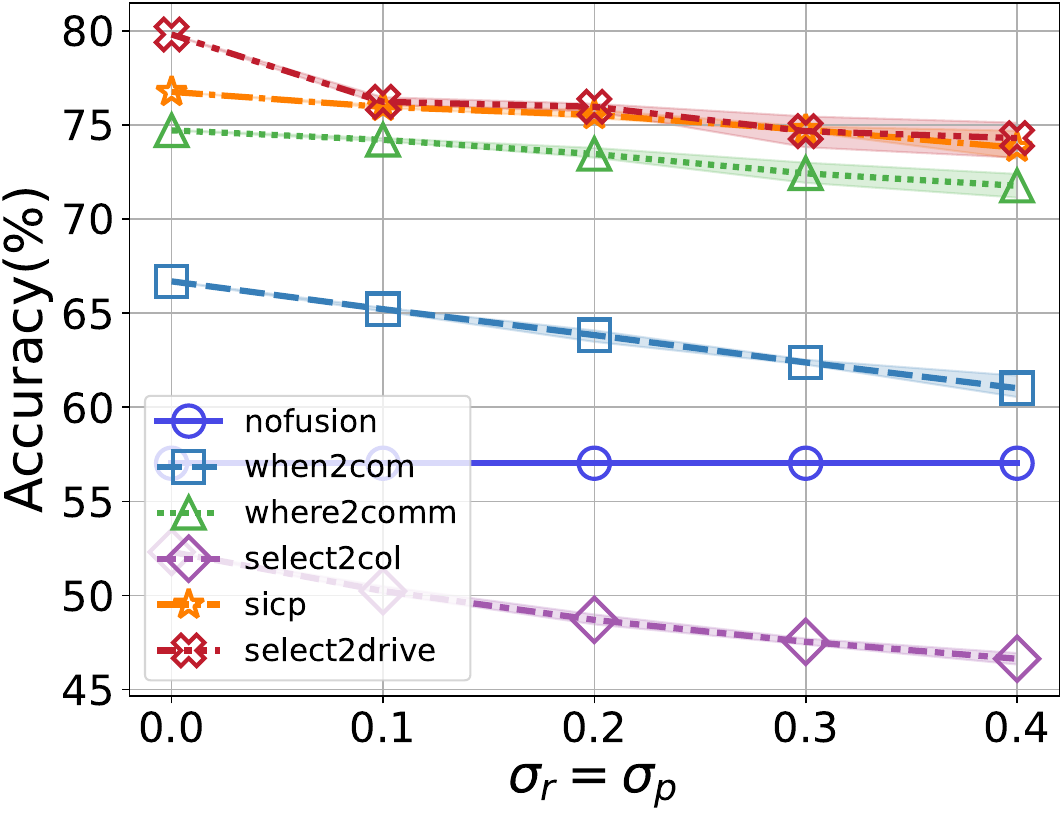}} 
    \caption{Robustness of DPP to the vehicle pose noise in the perception task, where uniform latency $\LatencyTrans$ is set to $100$ ms.}
    \label{Fig.noise}
    \vspace{-4mm}
\end{figure*}

\begin{figure}[tbp]
	\centering
    \subfloat[AP for $\LatencyAsync$ in $U(-100, 100)$ ms in V2XVerse.]{\includegraphics[width=0.5\linewidth]{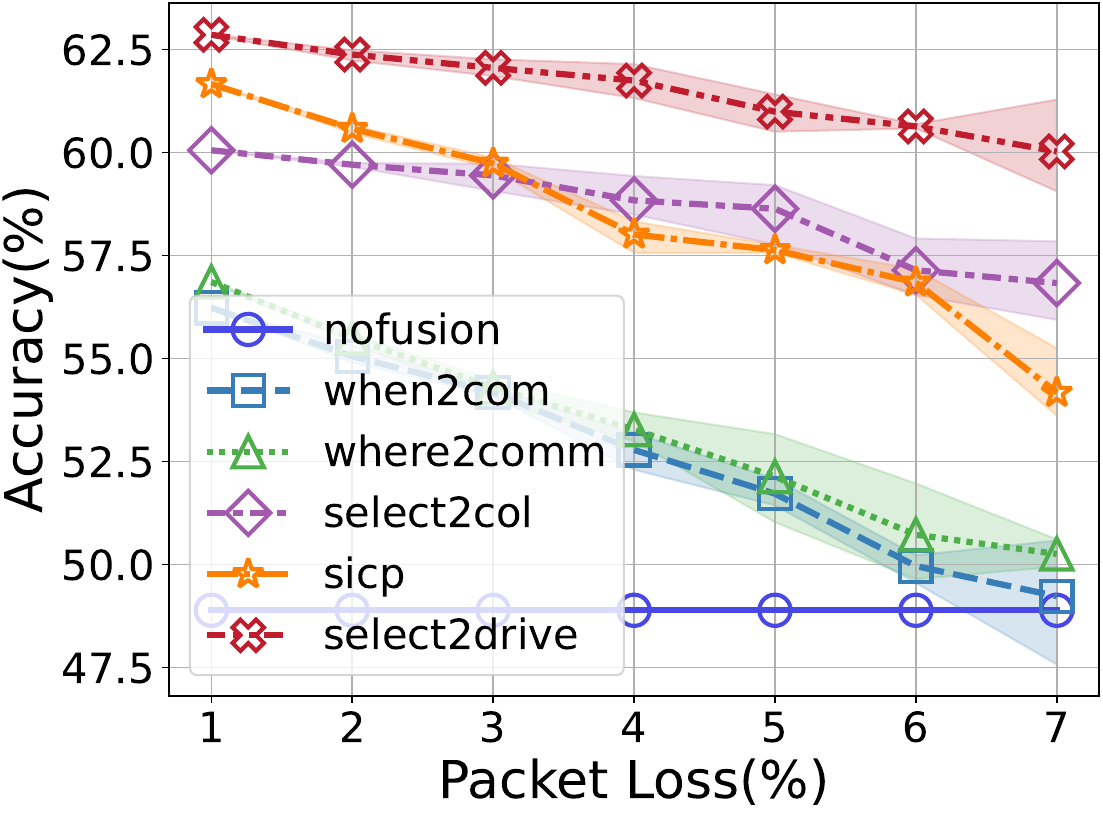}} 
    \subfloat[AP for $\LatencyAsync$ in $U(-50, 50)$ ms in V2XVerse.]{\includegraphics[width=0.5\linewidth]{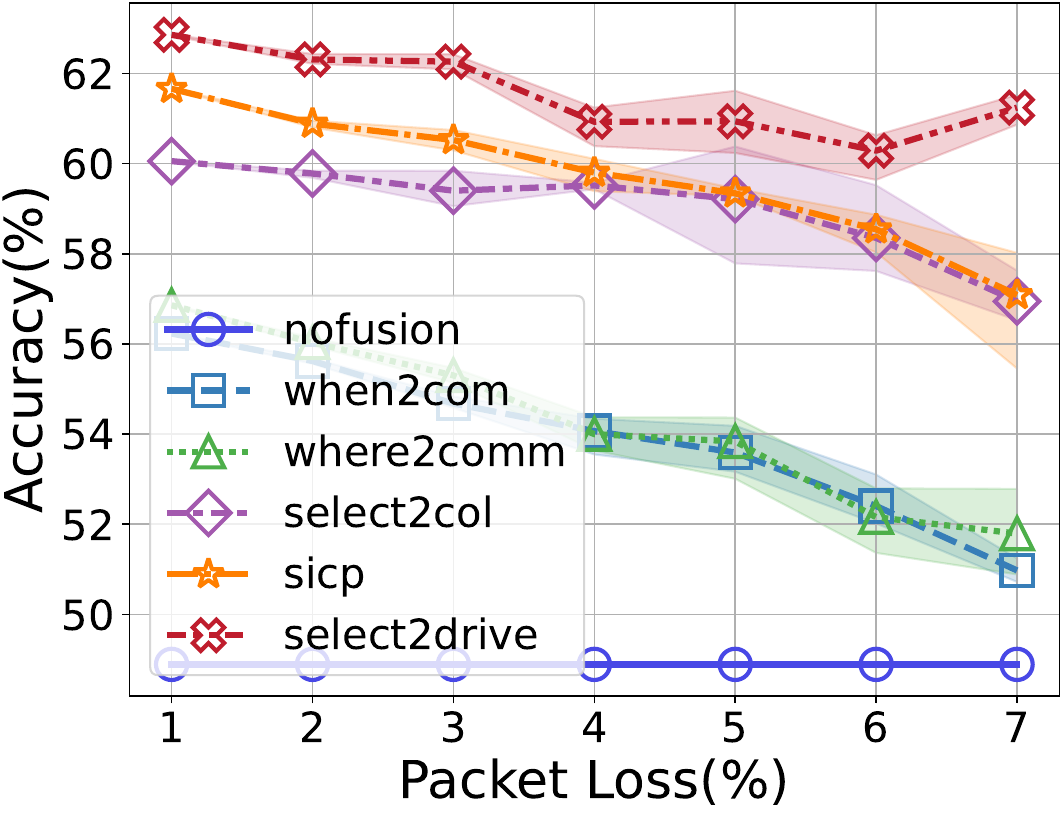}} \\
    \subfloat[AP for $\LatencyAsync$ in $U(-100, 100)$ ms in DAIR-V2X.]{\includegraphics[width=0.5\linewidth]{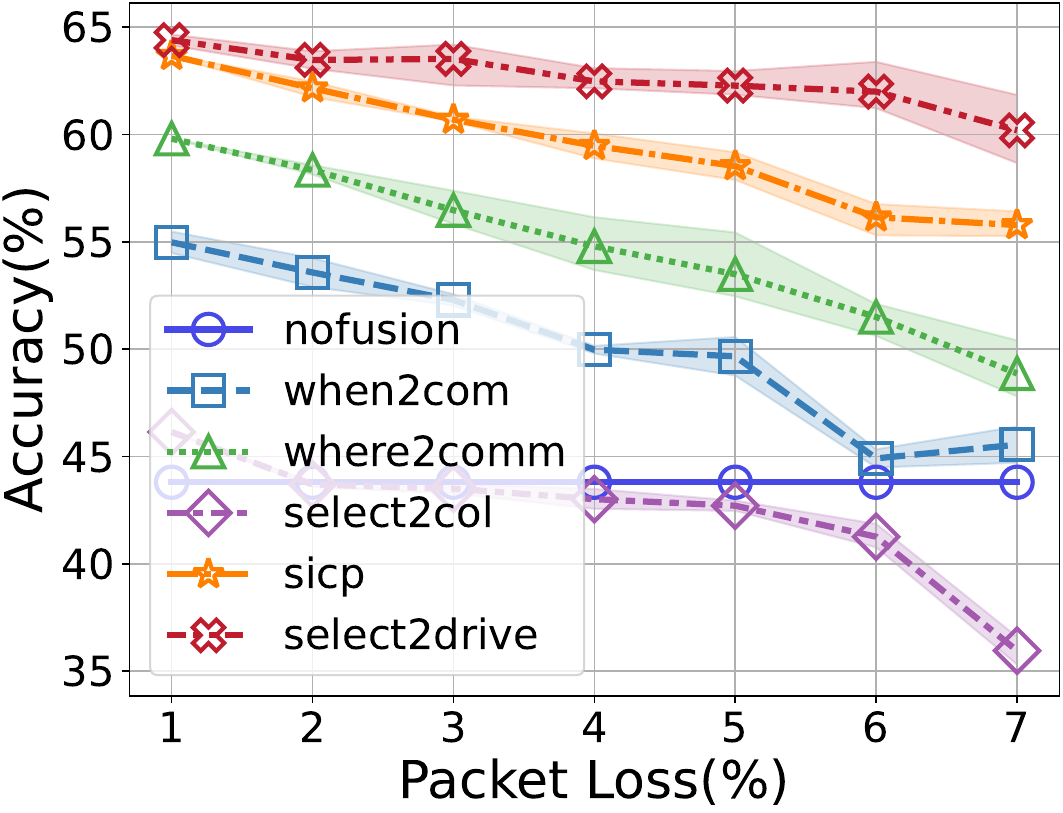}} 
    \subfloat[AP for $\LatencyAsync$ in $U(-50, 50)$ ms in DAIR-V2X.]{\includegraphics[width=0.5\linewidth]{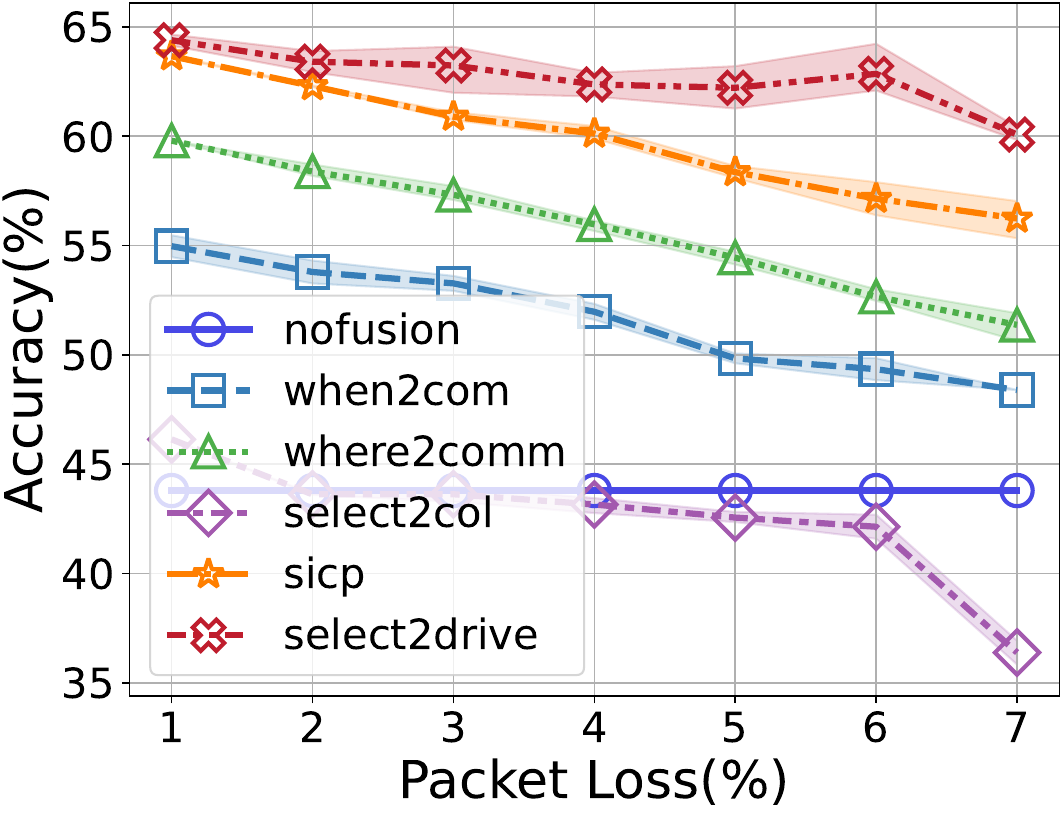}} 
    \caption{Robustness of DPP to the jitter and packet loss in the perception task, where uniform latency $\LatencyTrans$ is set to $100$ ms.}
    \label{Fig.packet}
    \vspace{-4mm}
\end{figure}

\subsubsection{Perception Performance}

Fig. \ref{Fig.visualize} presents qualitative findings. It can be observed that the predictive approach in Select2Drive facilitates the timely acquisition of projected data from surrounding vehicles, thus effectively addressing blind zone perception.
Due to the neglect of latency, Where2Comm employs outdated information aggregation, and consequently generates notable perceptual inaccuracies. Meanwhile, Select2Col and SiCP offer partial remediation, yet remain susceptible to blind zone perception loss stemming from temporal constraints. 
%

Fig. \ref{Fig.bandwidth} illustrates the perceptual performance across various methodologies under DSRC-based and C-V2X-based transmission scenarios. 
Under ideal communication conditions, a comprehensive multi-object perception evaluation indicates our method outperforms the other baselines. 
Meanwhile, in realistic V2X scenarios where existing methods degrade to performance levels comparable to non-communicative baselines, DPP maintains consistent performance advantages: achieving gains of {$2.60\%$} ($10$ MHz bandwidth)
and {$2.32\%$} ($300$ ms latency) 
over the second-best Select2Col method in V2Xverse, {while demonstrating improvements of $1.99\%$ ($10$ MHz) 
and $0.47\%$ ($300$ ms) 
against the second-best SiCP approach in DAIR-V2X.} 
Even under severely constrained bandwidth conditions, our method still maintains superior accuracy.
{The performance variation across different random seeds remains within $3\%$ for all primary methods, with the illustrated gains calculated as the average difference.}

As illustrated in Fig. \ref{Fig.noise}, DPP demonstrates robust stability in the presence of angular noise $\NoiseRotation$. 
The angular noise, parameterized by a standard deviation $\NoiseRotation$ in degrees, is formally modeled using a von Mises (or circular normal) distribution for the angle $\alpha$ with a concentration parameter $\kappa$ given by the relation $\kappa=(180/(\pi\cdot\NoiseRotation))^2$. 
For cases of low dispersion (i.e., large $\kappa$), this distribution is well-approximated by a normal distribution with a zero mean and a variance of $\sigma^2=(\pi\cdot\NoiseRotation/180)^2$. 
Meanwhile, its performance is slightly compromised when subjected to positioning noise $\NoisePosition$ (i.e., absolute deviations in $(x, y, z)$). 
Nevertheless, under moderate noise conditions, our method achieves significant gains of $3.27\%$ ($\NoisePosition = 0.1$, $\NoiseRotation=0$), 
$1.87\%$ ($\NoiseRotation=0.1$, $\NoiseRotation=0$), 
and $3.93\%$ ($\NoisePosition=\NoiseRotation=0.1$) 
compared to non-collaborative perception schemes in V2Xverse. {Meanwhile, the gain in DAIR-V2X is $19.55\%$ ($\NoisePosition = 0.1$, $\NoiseRotation=0$), 
$18.94\%$ ($\NoiseRotation = 0.1$, $\NoisePosition = 0$)
, and $14.84\%$ ($\NoisePosition=\NoiseRotation=0.1$).
}
Even under severe noise conditions, our method achieves significant gains of $1.79\%$ ($\NoisePosition = 0.2$, $\NoiseRotation=0$), 
$4.29\%$ ($\NoiseRotation=0.2$, $\NoisePosition = 0$), 
and $1.95\%$ ($\NoisePosition=\NoiseRotation=0.2$) 
compared to non-collaborative perception schemes in V2Xverse. {Meanwhile, the gain in DAIR-V2X is $17.98\%$ ($\NoisePosition = 0.2$, $\NoiseRotation = 0$), 
$17.48\%$ ($\NoiseRotation = 0.2$, $\NoisePosition = 0$), 
and $18.94\%$ ($\NoisePosition=\NoiseRotation=0.2$).}
This suggests the necessity of precise positional information, while our approach exhibits strong correction capabilities. 

{Fig. \ref{Fig.packet} represents the influence of different packet losses along with latency jitter on the performance. 
Specifically, when subjected to jitter with a variance of $100$ ms, DPP exhibits only a marginal precision reduction of less than $2\%$, while the second-best SiCP approach experiences a significant performance degradation of $6.6\%$ in the V2Xverse benchmark. 
Under elevated packet loss rates, our method exhibits reduced performance degradation and demonstrates superior robustness to burst jitter. 
This resilience is attributed to our DPP approach, which leverages temporal information from both preceding and succeeding frames. This multi-frame processing capability effectively mitigates communication failures caused by isolated spikes and enables cross-frame joint prediction to compensate for partial packet loss.}

\begin{figure}[tbp]
    \vspace{-1mm}
	\centering
        \subfloat[][Composited AP (solid line) and Communication Volume (dotted line) with different $\ThresholdCollab$]{\includegraphics[width=0.9\linewidth]{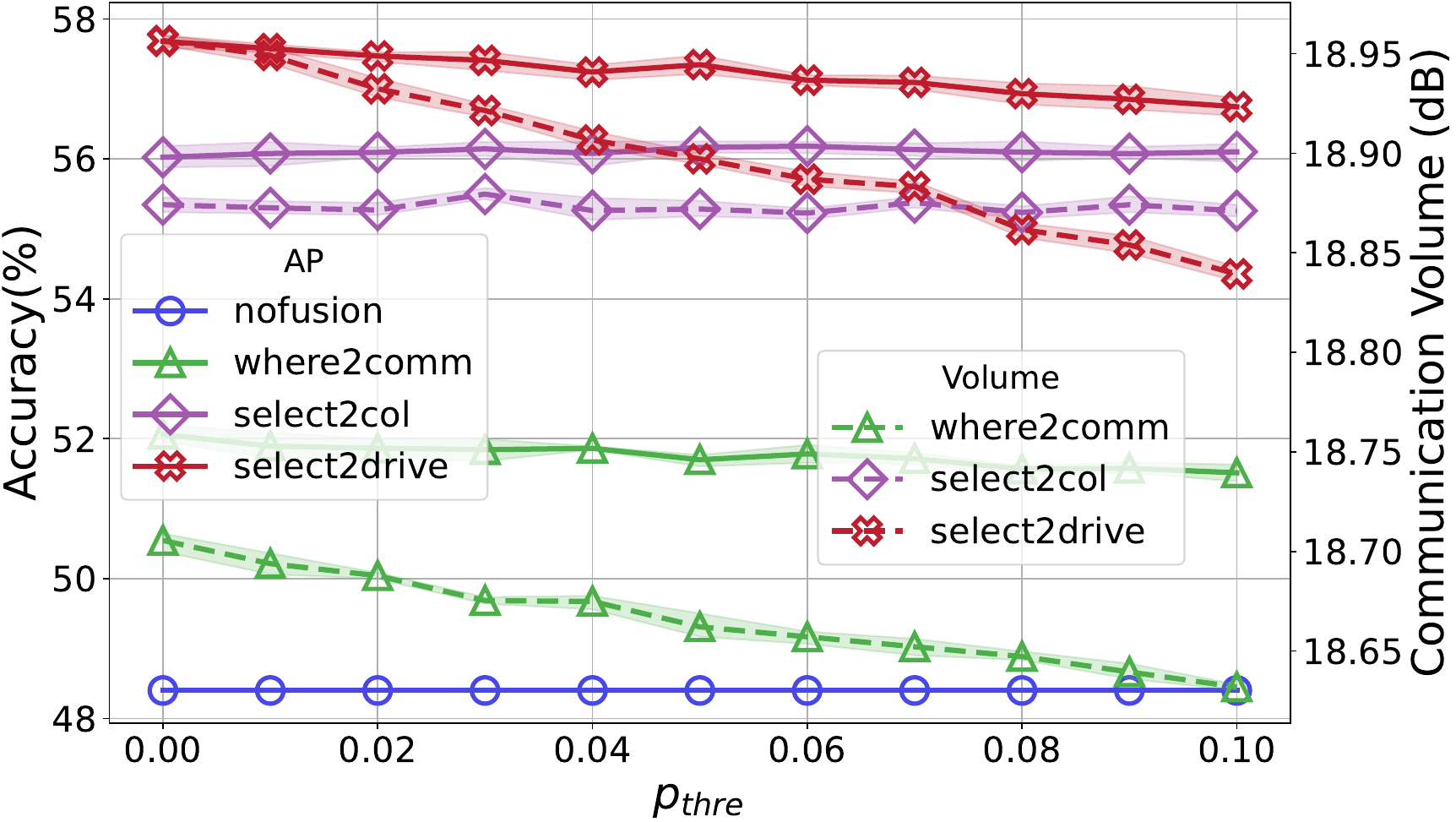}} 
        \hspace{0.2em}
        \subfloat[][ADE (solid line) and Communication Volume (dotted line) with different $\FocusRadius$]{\includegraphics[width=0.9\linewidth]{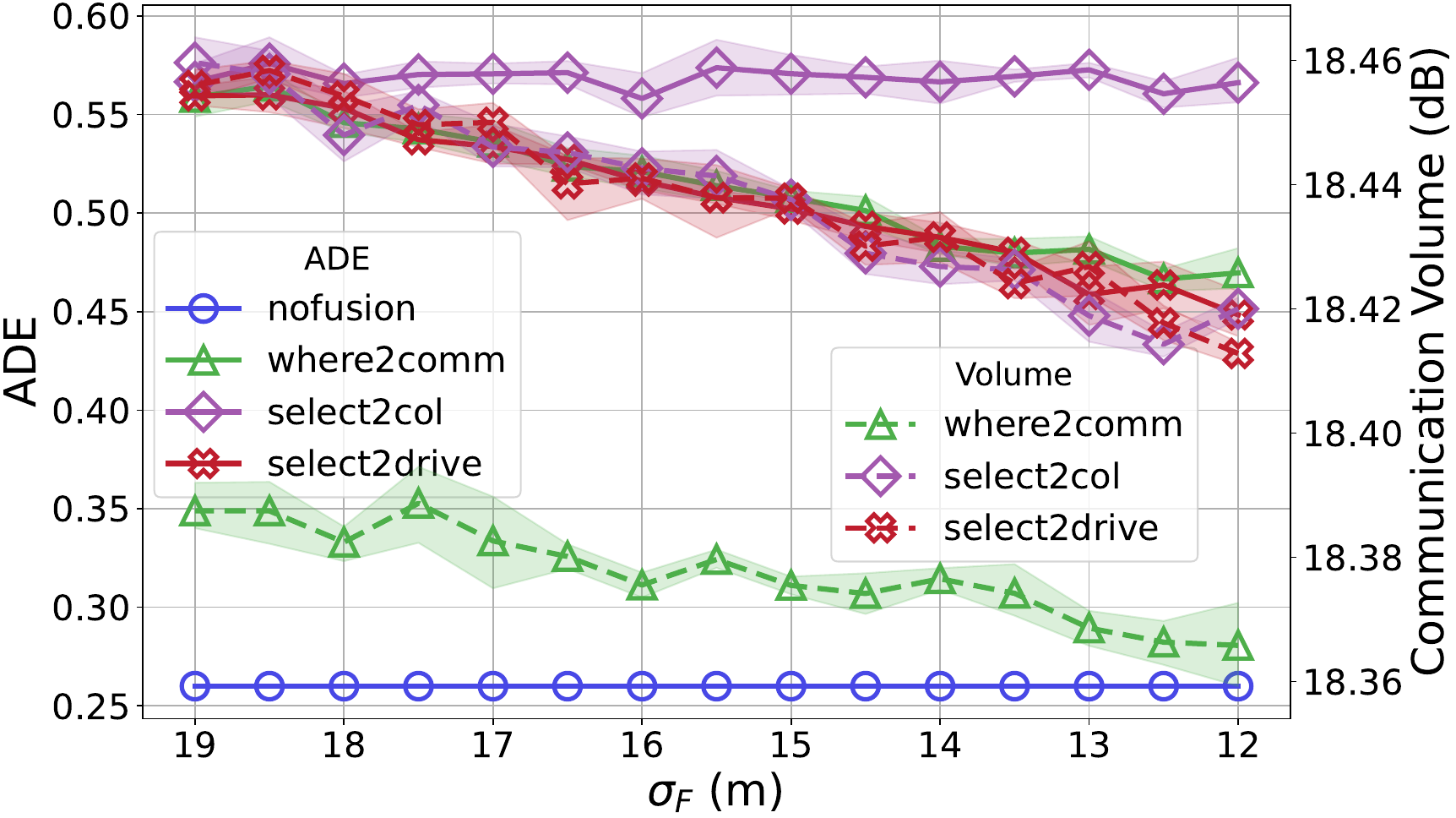}}
    \caption{Quantitative analysis on the influence of $\ThresholdCollab$ and $\FocusRadius$.}
    \label{Fig.hyperparameters}
    \vspace{-4mm}
\end{figure} 
\subsubsection{Hyperparameter Research} 
As depicted in Fig. \ref{Fig.hyperparameters}, we investigate the impact of the hyperparameter $\ThresholdCollab$ in Eq. \eqref{Eq.MessagePacking} and $\FocusRadius$ in Eq. \eqref{Eq.RequestMap}. 
The hyperparameter $\ThresholdCollab$ predominantly regulates the stringency of message exchange. 
A higher value diminishes the volume of information involved in aggregation, potentially inducing a degradation in perception performance. To achieve an optimal trade-off between communication efficiency and perception accuracy, we empirically set $\ThresholdCollab = 0.05$, referring to previous benchmarks \cite{xu2022v2x}. 
It can be observed from Fig. \ref{Fig.hyperparameters}(a), this configuration yields a decrease in communication overhead of $0.07$~dB ($4.74\%$),
accompanied by a marginal $0.33\%$ degradation in composite AP.
To determine the optimal $\FocusRadius$, we conduct an offline expert trajectory replication task. Performance is quantitatively assessed using the Average Displacement Error (ADE) \cite{alahi2016social}, 
which measures the mean Euclidean distance between the system's predicted trajectories and the ground truth expert demonstrations.
A lower value of $\FocusRadius$ enforces more stringent filtering of content extraneous to the driving task, thereby enhancing offline simulation performance during imitation of expert behaviors. However, under limited observational perspectives, the fidelity of expert imitation serves only as a reference metric rather than a direct determinant of ultimate performance, as the absence of collaboration leads to significant occlusions, illustrated in Fig. \ref{Fig.visualize}.
Consequently, Fig. \ref{Fig.hyperparameters}(b) indicates that setting $\FocusRadius = 15$ yielded an ADE reduction of $0.057$ ($10.20\%$), 
alongside a modest communication decrease of $0.023$ dB ($1.58\%$).

\normalsize
\section{Conclusions}\label{sec6}
In this work, we have presented Select2Drive, a PragComm-based real-time collaborative driving framework, which introduces two key components (i.e. DPP and APC) to address the critical timeliness challenges in V2X-AD systems.  
In particular, the DPP algorithm integrates predictive modeling and motion-aware affine transformation to infer future high-dimensional semantic features, maintaining robust perception performance even under severe positioning noise or constrained communication scenarios.  
Simultaneously, APC enhances decision-making efficiency by restricting communication to critical regions and minimizing unnecessary data exchanges, thereby mitigating potential confusion in decision-making.
Extensive evaluations have been conducted on both collaborative perception tasks and online closed-loop driving simulations. 
The experimental results demonstrate that our communication-efficient optimization framework is well-suited for real-time collaborative perception tasks, achieving significant performance improvements across a wide range of scenarios.
In the future, we will explore integrating generative models
to enhance driving policy robustness across diverse scenarios. 

\bibliographystyle{IEEEtran}
\vspace{-4mm}
\bibliography{reference} 

\begin{thebibliography}{10}
\providecommand{\url}[1]{#1}
\csname url@samestyle\endcsname
\providecommand{\newblock}{\relax}
\providecommand{\bibinfo}[2]{#2}
\providecommand{\BIBentrySTDinterwordspacing}{\spaceskip=0pt\relax}
\providecommand{\BIBentryALTinterwordstretchfactor}{4}
\providecommand{\BIBentryALTinterwordspacing}{\spaceskip=\fontdimen2\font plus
\BIBentryALTinterwordstretchfactor\fontdimen3\font minus \fontdimen4\font\relax}
\providecommand{\BIBforeignlanguage}[2]{{%
\expandafter\ifx\csname l@#1\endcsname\relax
\typeout{** WARNING: IEEEtran.bst: No hyphenation pattern has been}%
\typeout{** loaded for the language `#1'. Using the pattern for}%
\typeout{** the default language instead.}%
\else
\language=\csname l@#1\endcsname
\fi
#2}}
\providecommand{\BIBdecl}{\relax}
\BIBdecl

\bibitem{liu2023plant}
K.~Renz, K.~Chitta, O.-B. Mercea\emph{,~et~al.}, ``Plant: Explainable planning transformers via object-level representations,'' in \emph{Proceedings of the 6th Conference on Robot Learning}, Atlanta, GA, USA, 2023, pp. 459--470.

\bibitem{Shao2023ReasonNet}
H.~Shao, L.~Wang, R.~Chen\emph{,~et~al.}, ``Reasonnet: End-to-end driving with temporal and global reasoning,'' in \emph{2023 IEEE/CVF Conference on Computer Vision and Pattern Recognition (CVPR)}, Vancouver, BC, Canada, 2023, pp. 13\,723--13\,733.

\bibitem{peng2021learning}
Z.~Peng, Q.~Li, K.~M. Hui\emph{,~et~al.}, ``Learning to simulate self-driven particles system with coordinated policy optimization,'' in \emph{Advances in Neural Information Processing Systems}, vol.~34, 2021, pp. 10\,784--10\,797.

\bibitem{Sedar2023V2X}
R.~Sedar, C.~Kalalas, F.~Vázquez-Gallego\emph{,~et~al.}, ``A comprehensive survey of {V2X} cybersecurity mechanisms and future research paths,'' \emph{IEEE Open Journal of the Communications Society}, vol.~4, pp. 325--391, 2023.

\bibitem{xu2022opv2v}
R.~Xu, H.~Xiang, X.~Xia\emph{,~et~al.}, ``Opv2v: An open benchmark dataset and fusion pipeline for perception with vehicle-to-vehicle communication,'' in \emph{2022 International Conference on Robotics and Automation (ICRA)}, Philadelphia, PA, USA, 2022, pp. 2583--2589.

\bibitem{cui2022coopernaut}
J.~Cui, H.~Qiu, D.~Chen\emph{,~et~al.}, ``Coopernaut: End-to-end driving with cooperative perception for networked vehicles,'' in \emph{2022 IEEE/CVF Conference on Computer Vision and Pattern Recognition (CVPR)}, New Orleans, LA, USA, 2022, pp. 17\,231--17\,241.

\bibitem{wang2020v2vnet}
T.-H. Wang, S.~Manivasagam, M.~Liang\emph{,~et~al.}, ``{V2VNet}: Vehicle-to-vehicle communication for joint perception and prediction,'' in \emph{European Conference on Computer Vision}.\hskip 1em plus 0.5em minus 0.4em\relax Glasgow, UK: Springer, 2020, pp. 605--621.

\bibitem{xu2022v2x}
R.~Xu, H.~Xiang, Z.~Tu\emph{,~et~al.}, ``{V2X}-vit: Vehicle-to-everything cooperative perception with vision transformer,'' in \emph{European Conference on Computer Vision (ECCV)}, Tel Aviv, Israel, 2022, pp. 107--124.

\bibitem{3GPP_TS38.211}
\BIBentryALTinterwordspacing
{3rd Generation Partnership Project (3GPP)}, ``{Physical channels and modulation},'' 3GPP, Technical Specification TS 38.211 V18.2.0, Mar. 2024, (Release 18). [Online]. Available: \url{https://portal.3gpp.org/desktopmodules/Specifications/SpecificationDetails.aspx?specificationId=3180}
\BIBentrySTDinterwordspacing

\bibitem{hu2024pragmatic}
Y.~Hu, X.~Pang, X.~Qin\emph{,~et~al.}, ``Pragmatic communication in multi-agent collaborative perception,'' \emph{arXiv preprint arXiv:2401.12694}, 2024.

\bibitem{liu2024towards}
G.~Liu, Y.~Hu, C.~Xu\emph{,~et~al.}, ``Towards collaborative autonomous driving: Simulation platform and end-to-end system,'' \emph{IEEE Transactions on Pattern Analysis and Machine Intelligence}, 2025.

\bibitem{qu2024sicp}
D.~Qu, Q.~Chen, T.~Bai\emph{,~et~al.}, ``{SiCP}: Simultaneous individual and cooperative perception for {3D} object detection in connected and automated vehicles,'' in \emph{2024 IEEE/RSJ International Conference on Intelligent Robots and Systems (IROS)}, 2024, pp. 8905--8912.

\bibitem{liu2020when2com}
Y.-C. Liu, J.~Tian, N.~Glaser\emph{,~et~al.}, ``When2com: Multi-agent perception via communication graph grouping,'' in \emph{2020 IEEE/CVF Conference on Computer Vision and Pattern Recognition (CVPR)}, Seattle, WA, USA, 2020, pp. 4105--4114.

\bibitem{hu2022where2comm}
Y.~Hu, S.~Fang, Z.~Lei\emph{,~et~al.}, ``Where2comm: Communication-efficient collaborative perception via spatial confidence maps,'' in \emph{Advances in Neural Information Processing Systems (NeurIPS)}, New Orleans Convention Center, 2022, pp. 4874--4886.

\bibitem{Wei2023AsynchronyRobust}
S.~Wei, Y.~Wei, Y.~Hu\emph{,~et~al.}, ``Asynchrony-robust collaborative perception via bird's eye view flow,'' in \emph{Advances in Neural Information Processing Systems (NeurIPS)}, New Orleans, LA, USA, 2023, pp. 28\,462--28\,477.

\bibitem{lei2024robust}
Z.~Lei, Z.~Ni, R.~Han\emph{,~et~al.}, ``Robust collaborative perception without external localization and clock devices,'' in \emph{2024 IEEE International Conference on Robotics and Automation (ICRA)}, Yokohama, Japan, 2024, pp. 7280--7286.

\bibitem{liu2024select2col}
Y.~Liu, Q.~Huang, R.~Li\emph{,~et~al.}, ``Select2col: Leveraging spatial-temporal importance of semantic information for efficient collaborative perception,'' \emph{IEEE Transactions on Vehicular Technology}, vol.~73, no.~9, pp. 12\,556--12\,569, 2024.

\bibitem{10273599}
D.~Gündüz, F.~Chiariotti, K.~Huang\emph{,~et~al.}, ``Timely and massive communication in {6G}: Pragmatics, learning, and inference,'' \emph{IEEE BITS the Information Theory Magazine}, vol.~3, no.~1, pp. 27--40, 2023.

\bibitem{chen2020learning}
D.~Chen, B.~Zhou, V.~Koltun\emph{,~et~al.}, ``Learning by cheating,'' in \emph{Proceedings of the Conference on Robot Learning (CoRL)}, Auckland, New Zealand, 2020, pp. 66--75.

\bibitem{sun2025revisiting}
C.~Sun, P.~He, R.~Wang\emph{,~et~al.}, ``Revisiting communication efficiency in multi-agent reinforcement learning from the dimensional analysis perspective,'' arXiv preprint arXiv:2501.02888, 2025, [Online]. Available: http://arxiv.org/abs/2501.02888.

\bibitem{10258330}
P.~S. Chib and P.~Singh, ``Recent advancements in end-to-end autonomous driving using deep learning: A survey,'' \emph{IEEE Transactions on Intelligent Vehicles}, vol.~9, no.~1, pp. 103--118, 2024.

\bibitem{10.1145}
S.~So, J.~Petit, and D.~Starobinski, ``Physical layer plausibility checks for misbehavior detection in {V2X} networks,'' in \emph{Proceedings of the 12th Conference on Security and Privacy in Wireless and Mobile Networks}, New York, NY, USA, 2019, pp. 84--93.

\bibitem{liu2023maskma}
J.~Liu, Y.~Zhang, C.~Li\emph{,~et~al.}, ``{MaskMA}: Towards zero-shot multi-agent decision making with mask-based collaborative learning,'' arXiv preprint arXiv:2310.11846, 2023, [Online]. Available: http://arxiv.org/abs/2310.11846.

\bibitem{dosovitskiy2017carla}
A.~Dosovitskiy, G.~Ros, F.~Codevilla\emph{,~et~al.}, ``{CARLA}: An open urban driving simulator,'' in \emph{Proceedings of the 1st Annual Conference on Robot Learning}, Mountain View, CA, USA, 2017.

\bibitem{carlaleaderboard}
CARLA, ``Carla leaderboard,'' \url{https://leaderboard.carla.org/leaderboard/}.

\bibitem{shao2022interfuser}
H.~Shao, L.~Wang, R.~Chen\emph{,~et~al.}, ``Safety-enhanced autonomous driving using interpretable sensor fusion transformer,'' \emph{arXiv preprint arXiv:2207.14024}, 2022.

\bibitem{hao2025research}
R.~Hao, H.~Yu, J.~Zhong\emph{,~et~al.}, ``Research challenges and progress in the end-to-end {V2X} cooperative autonomous driving competition,'' arXiv preprint arXiv:2507.21610, 2025, [Online]. Available: http://arxiv.org/abs/2507.21610.

\bibitem{10229239}
C.~Zhang, F.~Steinhauser, G.~Hinz\emph{,~et~al.}, ``Occlusion-aware planning for autonomous driving with vehicle-to-everything communication,'' \emph{IEEE Transactions on Intelligent Vehicles}, vol.~9, no.~1, pp. 1229--1242, 2024.

\bibitem{ross2011reduction}
S.~Ross, G.~Gordon, and D.~Bagnell, ``A reduction of imitation learning and structured prediction to no-regret online learning,'' in \emph{Proceedings of the Fourteenth International Conference on Artificial Intelligence and Statistics}.\hskip 1em plus 0.5em minus 0.4em\relax Fort Lauderdale, FL, USA: JMLR Workshop and Conference Proceedings, 2011, pp. 627--635.

\bibitem{8460487}
F.~Codevilla, M.~Müller, A.~López\emph{,~et~al.}, ``End-to-end driving via conditional {Imitation Learning},'' in \emph{2018 IEEE International Conference on Robotics and Automation (ICRA)}, Brisbane, Australia, 2018, pp. 4693--4700.

\bibitem{9466501}
T.-Y. Tung, S.~Kobus, J.~P. Roig\emph{,~et~al.}, ``{Effective Communications}: A joint learning and communication framework for multi-agent reinforcement learning over noisy channels,'' \emph{IEEE Journal on Selected Areas in Communications}, vol.~39, no.~8, pp. 2590--2603, 2021.

\bibitem{gimenez2024semantic}
J.~M. Gim{\'e}nez-Guzm{\'a}n, I.~Leyva-Mayorga, and P.~Popovski, ``Semantic {V2X} communications for image transmission in {6G} systems,'' \emph{IEEE Network}, 2024.

\bibitem{IEEE802}
D.~Jiang and L.~Delgrossi, ``{IEEE 802.11p}: Towards an international standard for wireless access in vehicular environments,'' in \emph{2008 IEEE Vehicular Technology Conference (VTC Spring)}, Singapore, 2008, pp. 2036--2040.

\bibitem{7992934}
S.~Chen, J.~Hu, Y.~Shi\emph{,~et~al.}, ``Vehicle-to-everything (v2x) services supported by lte-based systems and 5g,'' \emph{IEEE communications standards magazine}, vol.~1, no.~2, pp. 70--76, 2017.

\bibitem{kenney2011dedicated}
J.~B. Kenney, ``Dedicated short-range communications ({DSRC}) standards in the united states,'' \emph{Proceedings of the IEEE}, vol.~99, no.~7, pp. 1162--1182, 2011.

\bibitem{9212349}
D.~Garcia-Roger, E.~E. González, D.~Martín-Sacristán\emph{,~et~al.}, ``{V2X} support in {3GPP} specifications: From {4G} to {5G} and beyond,'' \emph{IEEE Access}, vol.~8, pp. 190\,946--190\,963, 2020.

\bibitem{lei2022latency}
Z.~Lei, S.~Ren, Y.~Hu\emph{,~et~al.}, ``Latency-aware collaborative perception,'' in \emph{European Conference on Computer Vision (ECCV)}, Tel Aviv, Israel, 2022, pp. 316--332.

\bibitem{8356112}
Y.~Yao, B.~Xiao, G.~Wu\emph{,~et~al.}, ``{Multi-Channel} based {Sybil Attack Detection} in {Vehicular Ad Hoc Networks} using {RSSI},'' \emph{IEEE Transactions on Mobile Computing}, vol.~18, no.~2, pp. 362--375, 2019.

\bibitem{3GPP}
Q.~Zhu, C.-X. Wang, B.~Hua\emph{,~et~al.}, \emph{{3GPP} TR 38.901 Channel Model}.\hskip 1em plus 0.5em minus 0.4em\relax Wiley Press, 2021, pp. 1--35.

\bibitem{Gya2021V2X}
S.~Gyawali, S.~Xu, Y.~Qian\emph{,~et~al.}, ``Challenges and solutions for cellular based {V2X} communications,'' \emph{IEEE Communications Surveys and Tutorials}, vol.~23, no.~1, pp. 222--255, 2021.

\bibitem{Ashish2017Attention}
A.~Vaswani, N.~Shazeer, N.~Parmar\emph{,~et~al.}, ``Attention is all you need,'' in \emph{Advances in Neural Information Processing Systems (NeurIPS)}, Long Beach, CA, USA, 2017, pp. 5998--6008.

\bibitem{NMS}
A.~Neubeck and L.~Van~Gool, ``Efficient non-maximum suppression,'' in \emph{18th International Conference on Pattern Recognition (ICPR'06)}, vol.~3, Hong Kong, China, 2006, pp. 850--855.

\bibitem{jia2024bench2drive}
X.~Jia, Z.~Yang, Q.~Li\emph{,~et~al.}, ``Bench2drive: Towards multi-ability benchmarking of closed-loop end-to-end autonomous driving,'' \emph{arXiv preprint arXiv:2406.03877}, 2024.

\bibitem{wu2022tcp}
\BIBentryALTinterwordspacing
P.~Wu, X.~Jia, L.~Chen\emph{,~et~al.}, ``Trajectory-guided control prediction for end-to-end autonomous driving: A simple yet strong baseline,'' in \emph{Advances in Neural Information Processing Systems}, S.~Koyejo, S.~Mohamed, A.~Agarwal\emph{,~et~al.}, Eds., vol.~35.\hskip 1em plus 0.5em minus 0.4em\relax New Orleans, LA, USA: Curran Associates, Inc., 2022, pp. 6119--6132. [Online]. Available: \url{https://proceedings.neurips.cc/paper_files/paper/2022/file/286a371d8a0a559281f682f8fbf89834-Paper-Conference.pdf}
\BIBentrySTDinterwordspacing

\bibitem{SCOUT}
I.~Kotseruba and J.~K. Tsotsos, ``Understanding and modeling the effects of task and context on drivers’ gaze allocation,'' in \emph{2024 IEEE Intelligent Vehicles Symposium ({IV})}, Jeju, South Korea, 2024, pp. 1337--1344.

\bibitem{NEURIPS2023_6ca5d266}
H.~Yu, Y.~Tang, E.~Xie\emph{,~et~al.}, ``Flow-based feature fusion for vehicle-infrastructure cooperative 3d object detection,'' in \emph{Advances in Neural Information Processing Systems (NeurIPS)}, New Orleans, LA, USA, 2023, pp. 34\,493--34\,503.

\bibitem{2023DMVFN}
X.~Hu, Z.~Huang, A.~Huang\emph{,~et~al.}, ``A dynamic multi-scale voxel flow network for video prediction,'' in \emph{2023 IEEE/CVF Conference on Computer Vision and Pattern Recognition (CVPR)}, Vancouver, BC, Canada, 2023, pp. 6121--6131.

\bibitem{2017PredRNN}
Y.~Wang, Z.~Gao, M.~Long\emph{,~et~al.}, ``Predrnn++: Towards a resolution of the deep-in-time dilemma in spatiotemporal predictive learning,'' in \emph{Proceedings of the 35th International Conference on Machine Learning}, Stockholm, Sweden, 2018, pp. 5123--5132.

\bibitem{2023TAU}
C.~Tan, Z.~Gao, L.~Wu\emph{,~et~al.}, ``Temporal attention unit: Towards efficient spatiotemporal predictive learning,'' in \emph{2023 IEEE/CVF Conference on Computer Vision and Pattern Recognition (CVPR)}, Vancouver, BC, Canada, 2023, pp. 18\,770--18\,782.

\bibitem{2021MAU}
\BIBentryALTinterwordspacing
Z.~Chang, X.~Zhang, S.~Wang\emph{,~et~al.}, ``Mau: A motion-aware unit for video prediction and beyond,'' in \emph{Advances in Neural Information Processing Systems}, vol.~34, Virtual, 2021, pp. 26\,950--26\,962. [Online]. Available: \url{https://proceedings.neurips.cc/paper_files/paper/2021/file/e25cfa90f04351958216f97e3efdabe9-Paper.pdf}
\BIBentrySTDinterwordspacing

\bibitem{2020PhyDNet}
V.~L. Guen and N.~Thome, ``Disentangling physical dynamics from unknown factors for unsupervised video prediction,'' in \emph{2020 IEEE/CVF Conference on Computer Vision and Pattern Recognition (CVPR)}, Seattle, WA, USA, 2020, pp. 11\,471--11\,481.

\bibitem{lang2019pointpillars}
A.~H. Lang, S.~Vora, H.~Caesar\emph{,~et~al.}, ``Pointpillars: Fast encoders for object detection from point clouds,'' in \emph{2019 IEEE/CVF Conference on Computer Vision and Pattern Recognition (CVPR)}, Long Beach, CA, USA, 2019, pp. 12\,689--12\,697.

\bibitem{yin2021center}
T.~Yin, X.~Zhou, and P.~Kr{\"a}henb{\"u}hl, ``Center-based {3D} object detection and tracking,'' in \emph{Proceedings of the IEEE/CVF Conference on Computer Vision and Pattern Recognition}, 2021, pp. 11\,784--11\,793.

\bibitem{wu2020motionnet}
P.~Wu, S.~Chen, and D.~N. Metaxas, ``Motionnet: Joint perception and motion prediction for autonomous driving based on bird’s eye view maps,'' in \emph{2020 IEEE/CVF Conference on Computer Vision and Pattern Recognition (CVPR)}, Seattle, WA, USA, 2020, pp. 11\,382--11\,392.

\bibitem{NVIDIATHOR}
\BIBentryALTinterwordspacing
{NVIDIA}. (2022) {NVIDIA DRIVE Thor}. Accessed: June 16, 2025. [Online]. Available: \url{https://blogs.nvidia.com/blog/drive-thor/}
\BIBentrySTDinterwordspacing

\bibitem{8237740}
Z.~Liu, R.~A. Yeh, X.~Tang\emph{,~et~al.}, ``Video frame synthesis using deep voxel flow,'' in \emph{2017 IEEE International Conference on Computer Vision (ICCV)}, Venice, Italy, 2017, pp. 4473--4481.

\bibitem{NIPS2015_33ceb07b}
\BIBentryALTinterwordspacing
M.~Jaderberg, K.~Simonyan, A.~Zisserman\emph{,~et~al.}, ``Spatial transformer networks,'' in \emph{Advances in Neural Information Processing Systems}, vol.~28, Montreal, Canada, 2015. [Online]. Available: \url{https://proceedings.neurips.cc/paper_files/paper/2015/file/33ceb07bf4eeb3da587e268d663aba1a-Paper.pdf}
\BIBentrySTDinterwordspacing

\bibitem{jang2022categorical}
E.~Jang, S.~Gu, and B.~Poole, ``Categorical reparameterization with gumbel-softmax,'' in \emph{International Conference on Learning Representations}, Virtual, 2022.

\bibitem{bengio2013estimating}
Y.~Bengio, N.~Léonard, and A.~Courville, ``Estimating or propagating gradients through stochastic neurons for conditional computation,'' \emph{arXiv e-prints}, pp. arXiv--1308, 2013.

\bibitem{simonyan2015very}
K.~Simonyan and A.~Zisserman, ``Very deep convolutional networks for large-scale image recognition,'' in \emph{3rd International Conference on Learning Representations (ICLR)}, San Diego, CA, USA, 2015.

\bibitem{Jaeger2023ICCV}
B.~Jaeger, K.~Chitta, and A.~Geiger, ``Hidden biases of end-to-end driving models,'' in \emph{2023 IEEE/CVF International Conference on Computer Vision (ICCV)}, Paris Convention Center, 2023.

\bibitem{hu2020collaborative}
Y.~Hu, S.~Chen, Y.~Zhang\emph{,~et~al.}, ``Collaborative motion prediction via neural motion message passing,'' in \emph{2020 IEEE/CVF Conference on Computer Vision and Pattern Recognition (CVPR)}, Seattle, WA, USA, 2020, pp. 6318--6327.

\bibitem{OpenDrive}
ASAM, ``Asam opendrive standard,'' \url{https://www.asam.net/standards/detail/opendrive/}.

\bibitem{hwang2020communication}
R.-H. Hwang, M.~M. Islam, M.~A. Tanvir\emph{,~et~al.}, ``Communication and computation offloading for {5G} {V2X}: Modeling and optimization,'' in \emph{{GLOBECOM} 2020 -- 2020 {IEEE} Global Communications Conference}, Taipei, China, 2020, pp. 1--6.

\bibitem{wang2024motion}
Y.~Wang, H.~Chen, G.~Yin\emph{,~et~al.}, ``Motion state estimation of preceding vehicles with packet loss and unknown model parameters,'' \emph{IEEE/ASME Transactions on Mechatronics}, vol.~29, no.~5, pp. 3461--3472, 2024.

\bibitem{yu2022dair}
H.~Yu, Y.~Luo, M.~Shu\emph{,~et~al.}, ``{DAIR-V2X}: A large-scale dataset for vehicle-infrastructure cooperative {3D} object detection,'' in \emph{Proceedings of the IEEE/CVF Conference on Computer Vision and Pattern Recognition}, 2022, pp. 21\,361--21\,370.

\bibitem{alahi2016social}
A.~Alahi, K.~Goel, V.~Ramanathan\emph{,~et~al.}, ``Social {LSTM}: Human trajectory prediction in crowded spaces,'' in \emph{Proceedings of the {IEEE} Conference on Computer Vision and Pattern Recognition}, Las Vegas, NV, USA, 2016, pp. 961--971.

\end{thebibliography}
\vspace{-4mm}

\vspace{-10mm}
\begin{IEEEbiography}[{\includegraphics[width=1in,height=1.25in,clip,keepaspectratio]{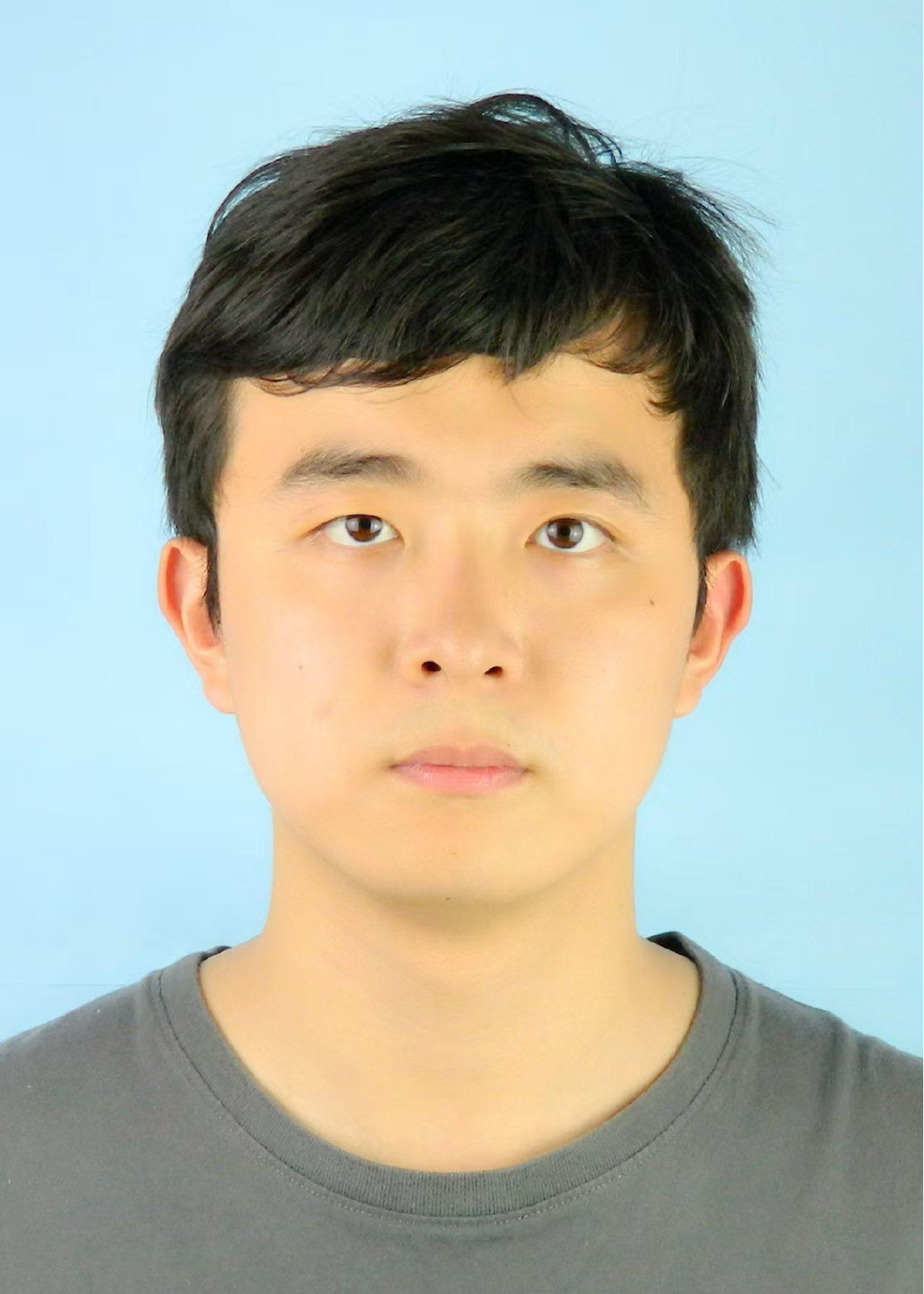}}]{Jiahao Huang}(Student Member, IEEE) received his B.E. in information engineering from Zhejiang University, Hangzhou, China, in 2023. He is currently pursuing a Ph.D with the College of Information Science and Electronic Engineering, Zhejiang University, Hangzhou, China. His research interests include autonomous driving, pragmatic communication, and deep reinforcement learning.
\end{IEEEbiography}

\begin{IEEEbiography}[{\includegraphics[width=1in,height=1.25in,clip,keepaspectratio]{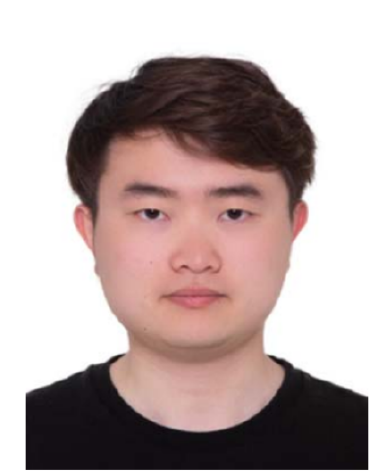}}]{Jianhang Zhu}(Graduate Student Member,IEEE) received the B.S. degree in communication engineering from Jilin University, Changchun, China, in 2020. He is currently working toward the EngD degree with the College of Information Science and Electronic Engineering, Zhejiang University, Hangzhou. His research interest includes graph neural network, multi-agent reinforcement learning, and edge computing.
\end{IEEEbiography}

\begin{IEEEbiography}[{\includegraphics[width=1in,height=1.25in,clip,keepaspectratio]{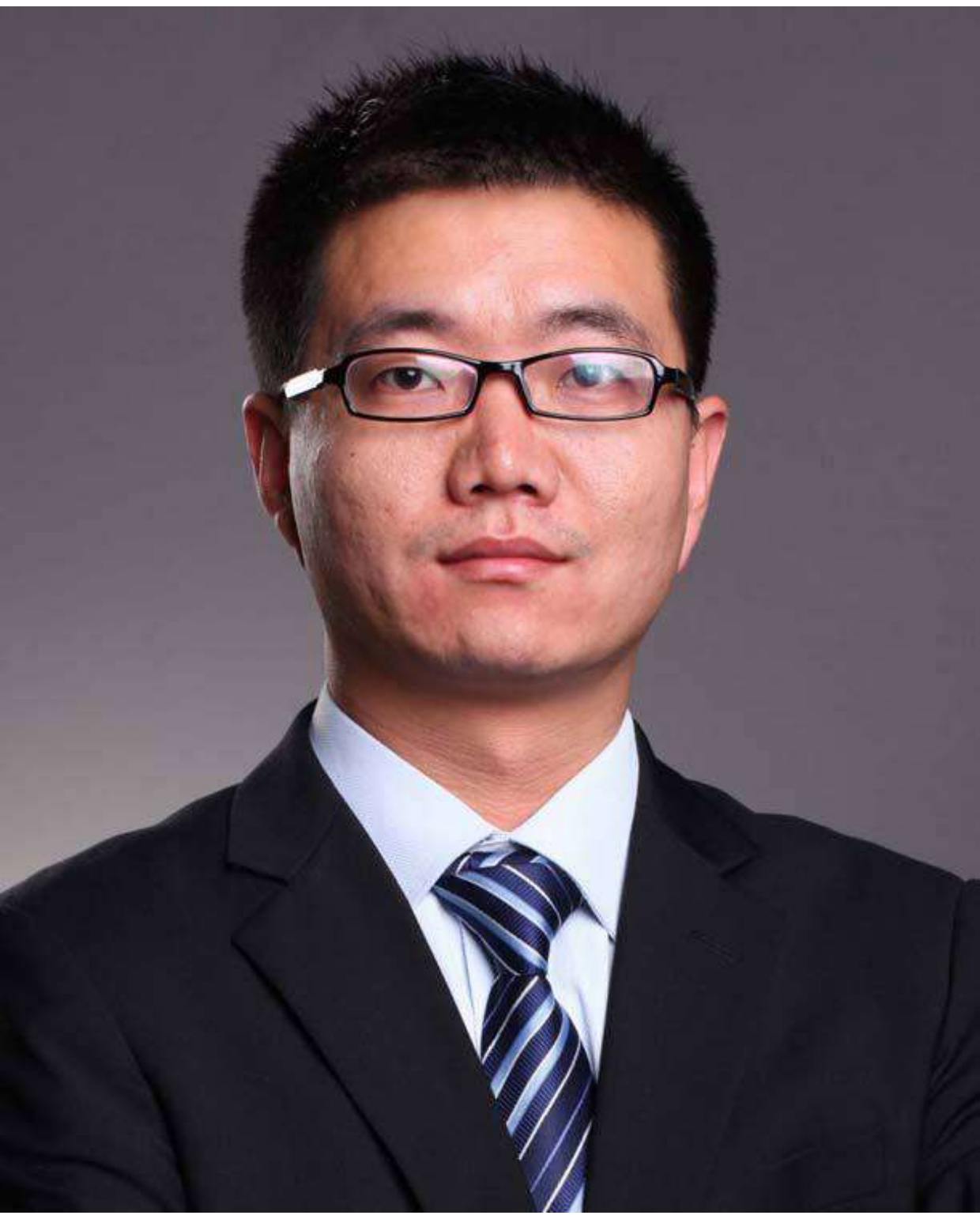}}]{Rongpeng Li}(Senior Member, IEEE) is currently an Associate Professor with the College of Information Science and Electronic Engineering, Zhejiang University. He received the B.E. degree from Xidian University, Xi’an, China, in June 2010, and the Ph.D. degree from Zhejiang University, Hangzhou, China, in June 2015. From August 2015 to September 2016, he was a Research Engineer with the Wireless Communication Laboratory, Huawei Technologies Company Ltd., Shanghai, China. He was a Visiting Scholar with the Department of Computer Science and Technology, University of Cambridge, Cambridge, U.K., from February 2020 to August 2020. His current research interests focus on networked intelligence for comprehensive efficiency (NICE).
\end{IEEEbiography}

\begin{IEEEbiography}[{\includegraphics[width=1in,height=1.25in,clip,keepaspectratio]{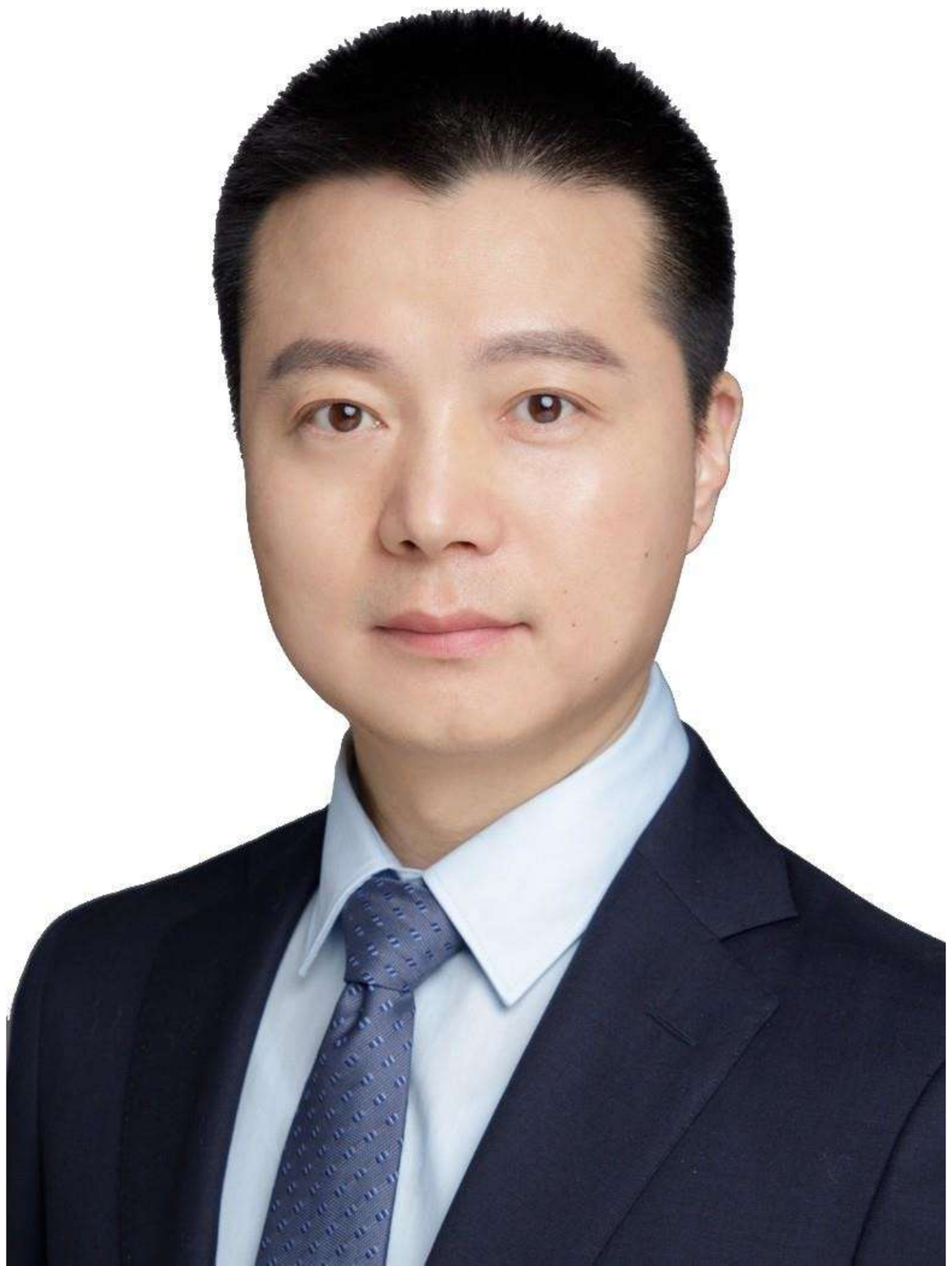}}]{Zhifeng Zhao}(Member, IEEE) received the B.E. degree in computer science, the M.E. degree in communication and information systems, and the Ph.D. degree in communication and information systems from the PLA University of Science and Technology, Nanjing, China, in 1996, 1999, and 2002, respectively. From 2002 to 2004, he acted as a Post-Doctoral Researcher with Zhejiang University, Hangzhou, China, where his researches were focused on multimedia next-generation networks (NGNs) and softswitch technology for energy efficiency. Currently, he is with the Zhejiang Lab, Hangzhou as the Chief Engineering Officer. His research areas include software defined networks (SDNs), wireless network in 6G, computing networks, and collective intelligence. He is the Symposium Co-Chair of ChinaCom 2009 and 2010. He is the Technical Program Committee (TPC) Co-Chair of the 10th IEEE International Symposium on Communication and Information Technology (ISCIT 2010).
\end{IEEEbiography}

\begin{IEEEbiography}[{\includegraphics[width=1in,height=1.25in,clip, keepaspectratio]{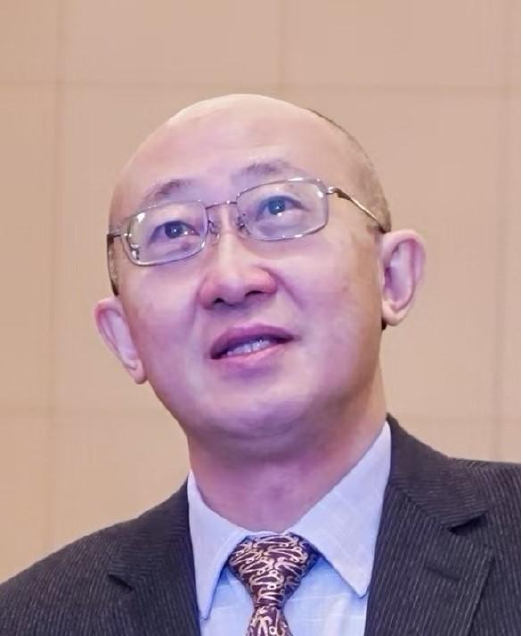}}]{Honggang Zhang}(Fellow, IEEE) was a Professor with the College of Information Science and Electronic Engineering, Zhejiang University, Hangzhou, China. He was an Honorary Visiting Professor with the University of York, York, U.K., and an International Chair Professor of Excellence with the Université Européenne de Bretagne, Supélec, France. He is a Professor with the School of Computer Science and Engineering, Macau University of Science and Technology, Macau, China. His research interests include cognitive radio networks, semantic communications, green communications, machine learning, artificial intelligence, intelligent computing, and Internet of Intelligence.

Dr. Zhang is a co-recipient of the 2021 IEEE Communications Society Outstanding Paper Award and the 2021 IEEE \textsc{INTERNET OF THINGS JOURNAL} (IOT-J) Best Paper Award. He served as the Chair of the Technical Committee on Cognitive Networks of the IEEE Communications Society from 2011 to 2012. He was the founding Chief Managing Editor of \textit{Intelligent Computing}, a Science Partner Journal. He was the leading Guest Editor for the Special Issues on Green Communications of the IEEE \textit{Communications Magazine}. He served as a Series Editor for the \textit{IEEE Communications Magazine} (Green Communications and Computing Networks Series) from 2015 to 2018. He is the Associate Editor-in-Chief of \textit{China Communications}.
\end{IEEEbiography}

\end{document}